\documentclass[apjl]{emulateapj}

\pdfoutput=1

\usepackage{comment}
\usepackage{ifthen}
\usepackage{amsmath}
\usepackage{amsfonts}
\usepackage{amssymb}
\usepackage{multirow}
\usepackage{wasysym}
\usepackage{subfigure}
\usepackage{epsfig}
\usepackage{wasysym}
\usepackage[dvipsnames]{xcolor}
\usepackage[%
	colorlinks=true,	% false: boxed links; true: colored links
	urlcolor=MidnightBlue,		% color of external links
	pdfpagelabels=true,
	hypertexnames=true,
	plainpages=false,
	naturalnames=true,
	pdftitle={A Transiting Jupiter Analog},    % title
	pdfauthor={Kipping et al.},     % author
	linkcolor=WildStrawberry,          % color of internal links (change box color with linkbordercolor)
	citecolor=ForestGreen,        % color of links to bibliography
        ]{hyperref}
\newcommand{\multi}{{\sc MultiNest}}
\newcommand{\cofiam}{{\tt CoFiAM}}
\newcommand{\blender}{{\tt BLENDER}}
\newcommand{\eccsamples}{{\tt ECCSAMPLES}}

\newcommand{\kepler}{\emph{Kepler}}
\newcommand{\Kepler}{\emph{Kepler}}
\newcommand{\wwwcoolworlds}{\href{https://github.com/CoolWorlds/Kepler-167-Posteriors}{this URL}}
\newcommand{\msun}{M_{\odot}}
\newcommand{\rsun}{R_{\odot}}
\newcommand{\mearth}{M_{\oplus}}
\newcommand{\rearth}{R_{\oplus}}
\newcommand{\mjup}{M_{\mathrm{J}}}
\newcommand{\rjup}{R_{\mathrm{J}}}
\newcommand{\lsun}{L_{\odot}}

\newcommand{\Teff}{T_{\mathrm{eff}}}

\def\kms{\ifmmode{\rm km\thinspace s^{-1}}\else km\thinspace s$^{-1}$\fi}
\def\ms{\ifmmode{\rm m\thinspace s^{-1}}\else m\thinspace s$^{-1}$\fi}
\newcommand{\thetastar}{ \boldsymbol{\theta_{\star}} }
\newcommand{\thetaplan}{ \boldsymbol{\theta_{\mathrm{P}}} }
\newcommand{\kic}{KIC-3239945}
\newcommand{\koi}{KOI-490}
\newcommand{\kep}{Kepler-167}
\newcommand{\koib}{KOI-490.01}
\newcommand{\kepb}{Kepler-167b}
\newcommand{\koic}{KOI-490.03}
\newcommand{\kepc}{Kepler-167c}
\newcommand{\koid}{KOI-490.04}
\newcommand{\kepd}{Kepler-167d}
\newcommand{\koie}{KOI-490.02}
\newcommand{\kepe}{Kepler-167e}

\newboolean{emulateapj}
\setboolean{emulateapj}{true}

\newboolean{astroph}
\setboolean{astroph}{true}

\newboolean{png}
\setboolean{png}{false}

\newboolean{color}
\setboolean{color}{true}

\shortauthors{Kipping et al.}
\shorttitle{A Transiting Jupiter Analog}
\ifthenelse{\boolean{emulateapj}}{
    
}{
    
}

\begin{document}

%% Titlepage
\title {A Transiting Jupiter Analog
%\altaffilmark{\titledag}
}

%% Authors
\author{
	{\bf	D.~M.~Kipping\altaffilmark{1},
		G.~Torres\altaffilmark{2},
		C.~Henze\altaffilmark{3},
		A.~Teachey\altaffilmark{1},
		H.~Isaacson\altaffilmark{4},\\
		E.~Petigura\altaffilmark{4},
		G.~W.~Marcy\altaffilmark{4},
		L.~A.~Buchhave\altaffilmark{5},
		J.~Chen\altaffilmark{1},
		S.~T.~Bryson\altaffilmark{3},
		E.~Sandford\altaffilmark{1}
	}
}

\altaffiltext{1}{Dept. of Astronomy, Columbia University, 550 W 120th St., New York, NY 10027, USA; email: dkipping@astro.columbia.edu}

\altaffiltext{2}{Harvard-Smithsonian Center for Astrophysics,
		Cambridge, MA 02138, USA}
		
\altaffiltext{3}{NASA Ames Research Center, Moffett Field, CA 94035, USA}

\altaffiltext{4}{University of California, Berkeley, CA 94720, USA}

\altaffiltext{5}{Centre for Star and Planet Formation, Natural History Museum of 
                 Denmark, University of Copenhagen, DK-1350 Copenhagen, Denmark}

%\altaffiltext{$\dagger$}{
%Based on archival data of the \emph{Kepler} telescope. 
%}

%% EOF authors

% #####################################################################
%% abstract
\begin{abstract}

Decadal-long radial velocity surveys have recently started to discover
analogs to the most influential planet of our solar system, Jupiter.
Detecting and characterizing these worlds is expected to shape our
understanding of our uniqueness in the cosmos. Despite the great successes
of recent transit surveys, Jupiter analogs represent a terra incognita,
owing to the strong intrinsic bias of this method against long orbital
periods. We here report on the first validated transiting
Jupiter analog, \kepe\ (\koie), discovered using \Kepler\ archival photometry
orbiting the K4-dwarf \kic. With a radius of $(0.91\pm0.02)$\,$\rjup$, a
low orbital eccentricity ($0.06_{-0.04}^{+0.10}$) and an equilibrium
temperature of $(131\pm3)$\,K, \kepe\ bears many of the basic hallmarks of
Jupiter. \kepe\ is accompanied by three Super-Earths on compact orbits, 
which we also validate, leaving a large cavity of transiting worlds around the
habitable-zone. With two transits and continuous photometric coverage, we are 
able to uniquely and precisely measure the orbital period of this post snow-line 
planet ($1071.2323\pm0.0006$\,d), paving the way for follow-up of this $K=11.8$
mag target.

\end{abstract}

% #####################################################################
%% keywords
\keywords{
	techniques: photometric --- planetary systems ---
        planets and satellites: detection --- stars: individual 
        (\kic, \koi, \kep)
}

%% EOF keywords
%% EOF titlepage

% #####################################################################
%% Introduction
\section{INTRODUCTION}
\label{sec:intro}

%% INTRODUCTION
%%

Jupiter is the dominant member of our planetary system with a mass exceeding twice 
that of all the other planets combined. Theories of the formation and evolution of
our neighboring planets are usually conditioned upon the properties and location
of our system's gargantuan world (see, e.g. \citealt{walsh:2011}), yet exoplanetary
surveys have only recently begun to assess the prevalence of such objects (see, e.g.
\citealt{gould:2010}).

Jupiter's presiding mass led to it playing a critical role in the dynamical evolution
of the early Solar System \citep{morbidelli:2007}. The final architecture of our
solar system, including the Earth, is thus intimately connected to the existence and
dynamical history of Jupiter \citep{batygin:2015}. The mass, location and existence
of Jupiter also likely affect the impact rate of minor bodies onto the Earth
\citep{horner:2010}, thereby influencing the evolution of terrestrial life.
The search for Jupiter analogs has therefore emerged as a scientific priority,
linked to the fundamental goal of understanding our uniqueness in the cosmos.

Around 20 extrasolar Jupiter analogs have been discovered with the radial velocity
method (see Table~4 of \citealt{rowan:2015}), indicating that these cool worlds are
not unique to the Solar System. Occurrence rate estimates, including constraints from
microlensing surveys, typically converge at $\eta_{\jupiter}\simeq3$\%
\citep{cumming:2008,gould:2010,wittenmyer:2011,rowan:2015} (depending upon the
definition of an ``analog''), although recently \citet{wittenmyer:2016} argued for
$6.1_{-1.6}^{+2.8}$\%. It is interesting to note that $\eta_{\jupiter}$ is approximately
equal to the prevalance of Earth analogs orbiting FGK stars as measured using the
\Kepler\ transit survey, specifically $\eta_{\oplus}=1.7_{-1.0}^{+2.6}$\%
\citep{petigura:2013,dfm:2014} (periods of 200--400\,days; 0.5--1.5\,$R_{\oplus}$),
although again with the caveat depending upon how one defines ``analogous''.

The transit method has dominated the exoplanet detection game over the last decade.
This technique has demonstrated a sensitivity to planets ranging from sub-Earths
\citep{barclay:2013} to super-Jupiters \citep{fortney:2011} orbiting a diverse array
of stars, such as M-dwarfs \citep{dressing:2015}, giants \citep{quinn:2015} and even
binaries \citep{doyle:2011}. From 2010-2015, the number of confirmed/validated
exoplanets discovered via the transit method is seven-fold that of all other exoplanet
hunting methods combined. One of the last regions of parameter-space which has been
stubbornly resistant to the reign of transits are those planets found beyond the
snow-line, owing to their long orbital periods. Indeed, whilst $\sim20$ Jupiter analogs
have been found with radial velocities \citep{rowan:2015}, no transiting examples have
been previously announced.

Jupiter analogs have both a reduced geometric transit probability and a lower chance
of being observed to transit within a fixed observing window less than twice
the planet's orbital period (which is usually the case). In general, one expects
the planey yield of a transit survey to scale as $P^{-5/3}$ \citep{beatty:2008},
implying that a 3\,day period Jupiter is $\sim$16,000 times easier to find than the
same planet at 1000\,days. Nevertheless, in a large transit survey spanning
multiple years, such as \Kepler\ ($\sim$200,000\,stars over $\sim$4\,years), these
obstacles are expected to yield and \Kepler\ should expect detections if the
occurrence rate is $\gtrsim\mathcal{O}[10^{-2}]$.

Pursuing this possibility, we here report the discovery of a 1071\,day period transiting
planet, \kepe\ (formeley \koie), orbiting the K-dwarf \kic\ (see
Table~\ref{tab:stellarproperties}) with a size and insolation comparable to Jupiter.
This planet, which lies comfortably beyond the snow-line, is found to transit twice over
the duration of the \Kepler\ mission allowing for a precise determination of the orbital
period and making it schedulable for future follow-up work. The data processing and 
follow-up observations required for this discovery are discussed in 
\S~\ref{sec:photometry}\,\&\,\S~\ref{sec:followup} respectively. In
\S~\ref{sec:validation}, we discuss how we are able to validate \kepe\ and the other
three transiting candidates (\koi.01, .03 \& .04; $P\sim4.4$, $7.4$ \& $21.8$\,d) within the
system using \blender. Light curve fits, leveraging asterodensity profiling, are discussed
in \S~\ref{sec:fits}, allowing us to infer the radius and even eccentricity of \kepe. Finally,
we place this discovery in context in \S~\ref{sec:discussion}, discussing the system
architecture and prospects for follow-up.

\section{KEPLER PHOTOMETRY}
\label{sec:photometry}

%% METHODS
%%

\subsection{Data Acquisition}
\label{sub:dataacquisition}

We downloaded the publicly available \kepler\ data for \koi\ from
the \href{http://archive.stsci.edu/}{Mikulski Archive for Space Telescopes}
(MAST). The downloaded data were released as part of Data Release 24 and were
processed using Science Operations Center (SOC) Pipeline version 9.2.24. All
quarters from 1-17 were available in long-cadence (LC) and from 9-17 there
was also short-cadence (SC), which was used preferentially over LC.

\subsection{Data Selection}
\label{sub:dataselection}

To fit light curve models to the \kepler\ data, it is necessary
to first remove instrumental and stellar photometric variability which can
distort the transit light curve shape. We break this process up into two stages:
(i) pre-detrending cleaning (ii) long-term detrending. In what follows, each 
quarter is detrended independently.

\subsection{Pre-detrending Cleaning}
\label{sub:cleaning}

The first step is to visually inspect each quarter and remove any exponential 
ramps, flare-like behaviors and instrumental discontinuities in the data. We 
make no attempt to correct these artifacts and simply exclude them from the 
photometry manually.

We inspect all points occurring outside of a transit for outliers. In-transit
points are defined as those occurring within $\pm0.6$ transit durations
of the nominal linear ephemeris for each KOI. For these durations
and ephemerides, we adopt the
\href{http://exoplanetarchive.ipac.caltech.edu}{NASA Exoplanet Archive} 
\cite{akeson:2013} parameters. We then
clean the out-of-transit Simple Aperture Photometry (SAP) light curve of
3\,$\sigma$ outliers, identified using a moving median smoothing curve
with a 20-point window.

\subsection{Detrending with \cofiam}
\label{sub:detrending}

For the data used in the transit light curve fits in \S\ref{sec:fits},
it is also necessary to remove the remaining long-term trends in the time
series. These trends can be due to instrumental effects, such as focus drift,
or stellar effects, such as rotational modulations. For this task, data are 
detrended using the Cosine Filtering with Autocorrelation Minimization (\cofiam) 
algorithm. \cofiam\ was specifically developed to protect the shape of a 
transit light curve and we direct the reader to our previous work 
\citep{hek:2013} for a detailed description.

Each transit of each KOI is detrended independently using \cofiam, setting the
protected timescale to twice the associated transit duration. After detrending,
the light curves were fitted with the same light curve model and algorithm
described later in \S\ref{sec:fits}. The maximum likelihood duration and
ephemeris were saved from these fits. We then used these values to go back to
\S\ref{sub:cleaning} and repeat the entire detrending process, to ensure we used
accurate estimates of these terms.

\section{FOLLOW-UP OBSERVATIONS}
\label{sec:followup}

%% FOLLOW-UP
%%

\subsection{Spectroscopy}
\label{sub:spectroscopy}

\koi\ was observed on 2011 October 16 at the Keck~I telescope on Mauna
Kea (HI) with the HIRES spectrometer \citep{vogt:1994}, in order to help
characterize the star as described below in \S~\ref{sub:stellar}.
The exposure time was 30 minutes and the spectrograph slit was set
using the C2 decker ($0\farcs86 \times 14\arcsec$). Reductions were
performed with the standard procedures employed by the California
Planet Search \citep{howard:2010,johnson:2010}. This resulted in an
extracted spectrum with $R \sim 60,000$ covering the approximate
wavelength range 360--800~nm, with a signal-to-noise ratio of 90 per
resolution element in the region of the \ion{Mg}{1}\,b triplet
(519~nm).

We examined the spectrum for signs of absorption lines from another
star that might be causing the transit signal, if located within the
slit. This was done by first subtracting a spectrum closely matching
that of the target star (after proper wavelength shifting and
continuum normalization), and then inspecting the residuals
\citep[see][]{kolbl:2015}. We saw no evidence of secondary spectral
lines. In order to quantify our sensitivity to such companions we
performed numerical simulations in which we subjected the residuals to
a similar fitting process by injecting mock companions over a range of
temperatures from 3500 to 6000~K, and with a broad range in relative
velocities. We then attempted to recover them, and this allowed us to
estimate that we are sensitive to companions down to about 1\% of the
flux of the primary star, with velocity separations greater than
10~\kms. For smaller relative velocities the secondary lines would be
blended with those of the primary and would not be detected. This
spectroscopic constraint is used below for the validation of the
candidates in \S~\ref{sec:validation}.

\subsection{High-resolution Imaging}
\label{sub:imaging}

Images from the $J$-band UK Infrared Telescope survey
\citep[UKIRT;][]{lawrence:2007} available on the \kepler\ Community
Follow-up Observing Program (CFOP) Web site\footnote{\tt
  https://cfop.ipac.caltech.edu/home/\,.} show a nearby companion
about five magnitudes fainter than \koi\ at an angular separation of
2\farcs1 in position angle 62\fdg7, which falls within the photometric
aperture of \kepler. Ancillary information for this source based on
automatic image classification indicates a probability of 99.4\% that
it is a galaxy, 0.3\% that it is a star, and 0.3\% that it is noise,
though it is unclear how robust these assessments are.  The UKIRT
images have a typical seeing-limited resolution of about 0\farcs8 or
0\farcs9. Additional imaging efforts reported on CFOP include speckle
interferometry observations on the WIYN 3.5m telescope at 692~nm and
880~nm \citep[see][]{horch:2014,everett:2015}, lucky imaging observations
on the Calar Alto 2m telescope at 766~nm \citep{lillo-box:2012}, and
imaging with the Robo-AO system on the Palomar 1.5m telescope
approximately in the $R$ band \citep{law:2014}, none of which detected
this companion likely due to its faintness at the optical wavelengths
probed by these observations.

To investigate this detection further and to explore the inner regions
around the target, we observed \koi\ with near-infrared adaptive
optics (AO) using the $1024 \times 1024$ NIRC2 imager
\citep{wizinowich:2004,johansson:2008} on the Keck\,II, 10m telescope on
the night of 2014 June 12. We used the natural guide star system, as
the target star was bright enough to be used as the guide star. The
data were acquired using the narrow-band Br-$\gamma$ filter and the
narrow camera field of view with a pixel scale of 9.942
mas~pixel$^{-1}$. The Br-$\gamma$ filter has a narrower bandwidth
(2.13--2.18~$\micron$), but a similar central wavelength
(2.15~$\micron$) compared the 2MASS $K_s$ filter
(1.95--2.34~$\micron$; 2.15~$\micron$) and allows for longer
integration times before saturation. A 3-point dither pattern was
utilized to avoid the noisier lower left quadrant of the NIRC2 array.
The 3-point dither pattern was observed three times with two coadds
per dither position for a total of 18 frames; each frame had an
exposure time of 30~s, yielding a total on-source exposure time of
$3\times 3\times 2\times 30~{\rm s} = 540$~s.  The target star was
measured with a resolution of 0\farcs051 (FWHM).

\begin{figure}
\epsscale{1.05}
\plotone{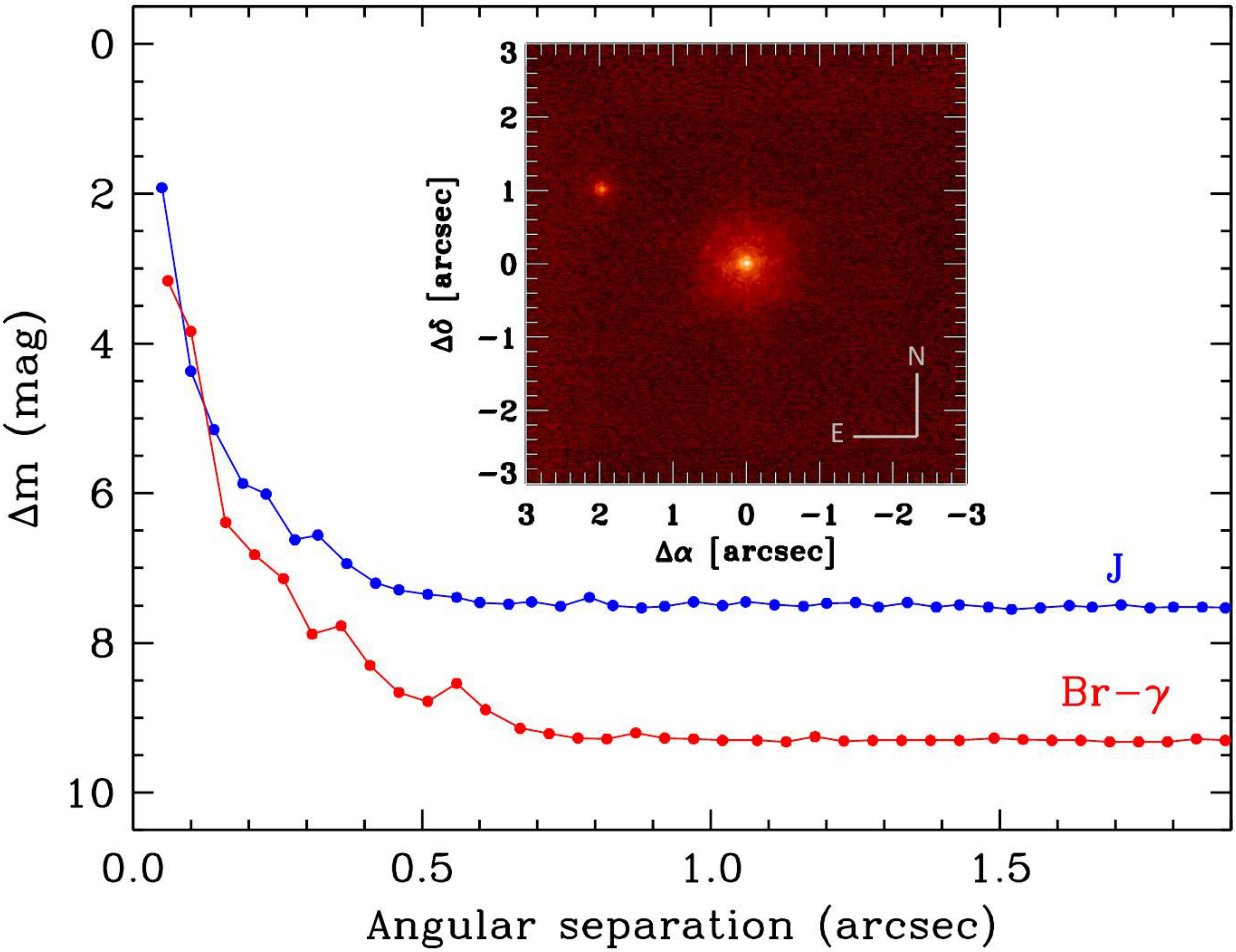}
\figcaption[]{
Br-$\gamma$ Keck/NIRC2 AO image of \koi\ shown along with the
sensitivity curves in the $J$ (1.248~$\micron$) and Br-$\gamma$
(2.15~$\micron$) bands.
\label{fig:ao}}
\end{figure}

The object 2\arcsec\ to the NE of \koi\ seen in the UKIRT images was
clearly detected in the NIRC2 data (see Figure~\ref{fig:ao}), and no
other stars were detected within 10\arcsec. The image of the companion
appears stellar, so we consider it a star in the following. In the
Br-$\gamma$ passband the data are sensitive to companions that have a
$K$-band contrast of $\Delta K = 3.8$~mag at a separation of 0\farcs1
and $\Delta K = 8.8$~mag at 0\farcs5 from the central star. We
estimated the sensitivities by injecting fake sources with a
signal-to-noise ratio of 5 into the final combined images at distances
of $N \times {\rm FWHM}$ from the central source, where $N$ is an
integer.

We also observed \koi\ in the $J$-band (1.248~$\micron$) with NIRC2 in
order to obtain the $J-K$ color of the companion star.  The $J$-band
observations employed the same 3-point dither pattern with an
integration time of 10~s per coadd for a total on-source integration
time of 180~s. These data had slightly better resolution (0\farcs046)
and a sensitivity of $\Delta J = 4.4$~mag at 0\farcs1 and $\Delta J =
7.4$~mag at 0\farcs5. Full sensitivity curves in both $J$ and
Br-$\gamma$ are shown in Figure~\ref{fig:ao}.  The companion star was
found to be fainter than the primary star by $\Delta J = 3.84\pm
0.03$~mag and $\Delta K = 3.51\pm 0.01$~mag, and separated from the
primary by $\Delta\alpha = 1\farcs97 \pm 0\farcs01$ and $\Delta\delta
= 1\farcs00 \pm 0\farcs01$ in right ascension and declination
(corresponding to an angular separation of 2\farcs21 in position angle
63\fdg1).  After deblending the 2MASS photometry we find that the
primary has $J$- and $K$-band magnitudes of $J_1 = 12.47 \pm 0.02$~mag
and $K_1 = 11.87 \pm 0.02$~mag, and the secondary has $J_2 = 16.32 \pm
0.04$~mag and $K_2 = 15.38 \pm 0.02$~mag. The companion is a redder
star than the primary: the individual colors are $(J-K)_1 = 0.60 \pm
0.03$ and $(J-K)_2 = 0.94 \pm 0.04$.\\

\subsection{Centroid motion analysis}
\label{sub:centroids}

The very precise astrometry that can be obtained from the
\kepler\ images enables a search for false positives that may be
causing one or more of the signals in \koi, such as a background
eclipsing binary. This can be done by measuring the location of the
transit signals relative to the target by means of difference images,
formed by subtracting an average of in-transit pixel values from
out-of-transit pixel values. If a transit signal is caused by a
stellar source, then the difference image will show that stellar
source, and its location can be determined by pixel response function
centroiding \citep{bryson:2013}. The centroid of an average
out-of-transit image provides the location of \koi\ because the object
is well isolated. The centroid of the difference image is then
compared to that of the out-of-transit image, which provides the
location of the transit source relative to \koi.

The automatic pipeline processing of \kepler\ provides these offsets
for each quarter in the Data Validation Reports, which are available
through the CFOP Web site. For \koib\ and \koic\ the multi-quarter
averages of the offsets indicate a position for the source of the
transits that is consistent with the location of target. Based on the
1$\sigma$ uncertainties associated with those multi-quarter average
offsets we adopted 3$\sigma$ radii of confusion for these two
candidates of 0\farcs396 and 0\farcs603, respectively, within which
the centroid motion analysis is insensitive to the presence of
contaminating stars. As these limits are smaller than the separation
of the 2\farcs2 companion reported earlier, that star cannot be the
source of these transits.

For \koie\ and \koid\ the automatic fits performed by the
\kepler\ pipeline failed, as indicated in the Data Validation Reports,
so no centroid information is available for these candidates.

\subsection{Stellar properties}
\label{sub:stellar}

The spectroscopic properties of \koi\ were determined from an analysis
of our Keck/HIRES spectrum. Our analysis was performed using the Stellar
Parameter Classification (SPC) pipeline \citep{buchhave:2012}, which
cross-correlates the observed spectrum against a large library of
calculated spectra based on model atmospheres by R.\ L.\ Kurucz, and
assigns stellar properties interpolating amongst those of the
synthetic spectra providing the best match. This analysis gave
$\Teff = (4890 \pm 50)$~K, $\log g = (4.61 \pm 0.10)$, ${\rm [Fe/H]} =
(-0.03 \pm 0.08)$, and $v \sin i < 2$~\kms.  The measured radial
velocity is $(-29.3 \pm 1.0)$~\kms, and the effective temperature
corresponds to a spectral type of K3 or K4.

\begin{figure*}
\begin{center}
\includegraphics[width=18.0 cm]{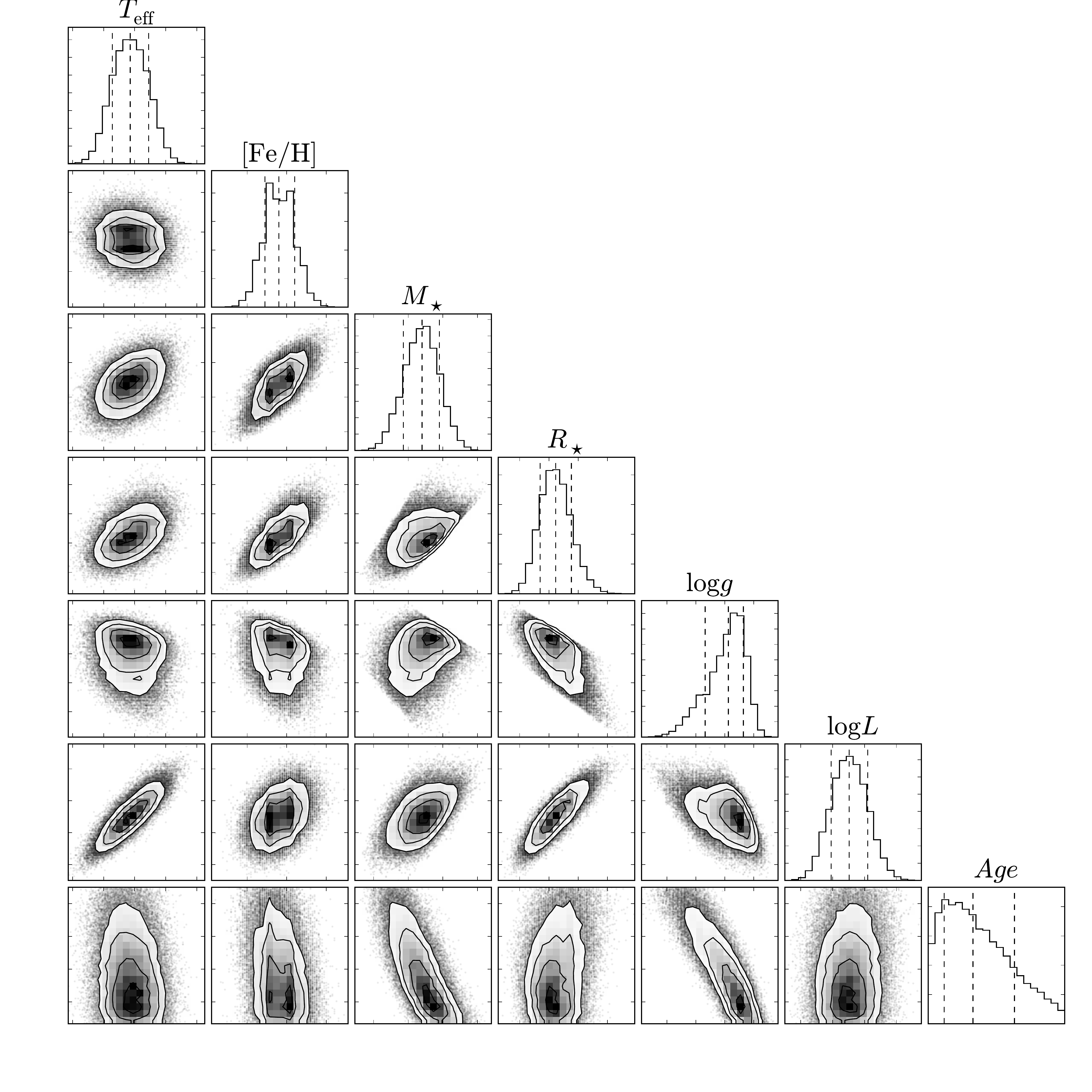}
\caption{
Triangle plot of the posterior distributions of fundamental stellar parameters
using SPC plus Dartmouth isochrones. Contours mark the 0.5, 1.0, 1.5 \&
2.0\,$\sigma$ confidence intervals and dashed lines on the histograms mark
the median and surrounding 1\,$\sigma$ confidence interval. Posteriors may
be downloaded at \wwwcoolworlds.
}
\label{fig:K167_post}
\end{center}
\end{figure*}

The mass and radius of the star, along with other properties, were
estimated by comparing the SPC parameters against a grid of Dartmouth
isochrones \citep{dotter:2008} with a $\chi^2$ procedure similar to 
that described by \cite{torres:2008}. Because the stellar radius and
age are largely determined by the surface gravity, and our $\log g$
determination provides a relatively weak constraint for \koi\ given 
its uncertainty, we supplemented it with an estimate of the mean 
stellar density obtained by fitting the \kepler\ light curves of 
\koib, \koic, and \koid, on the assumption that they are true 
planets (justified below) and that they orbit the same star. The 
stellar parameters derived in this way are listed in 
Table~\ref{tab:stellarproperties}, along with the inputs from SPC and 
the photometric mean density (posteriors shown in Figure~\ref{fig:K167_post}). 
The inferred distance is based on the apparent $K_s$-band magnitude from 2MASS
and a reddening estimate of $E(B-V) = (0.075 \pm 0.030)$ from the \kepler\ 
Input Catalog \citep[KIC;][]{brown:2011}.

\begin{deluxetable}{lc}
\tablewidth{0pc}
\tablecaption{Stellar properties of \koi.
\label{tab:stellarproperties}}
\tablehead{
\colhead{~~~~~~~~~~~Property~~~~~~~~~~~} & \colhead{Value}}
\startdata
$\Teff$ (K)\tablenotemark{a}\dotfill               & $4890 \pm 50$ \\ [+1pt]
$\log g$ (dex)\tablenotemark{a}\dotfill                  & $4.61 \pm 0.10$ \\ [+1pt]
${\rm [Fe/H]}$ (dex)\tablenotemark{a}\dotfill            & $-0.03 \pm 0.08$ \\ [+1pt]
$v \sin i$ (km~s$^{-1}$)\tablenotemark{a}\dotfill        & $< 2$ \\ [+1pt]
$\log_{10}[\rho_{\star}/($kg m$^{-3})]$\tablenotemark{b}\dotfill  & $3.460_{-0.065}^{+0.031}$ \\ [+2pt]
$M_{\star}$ ($\msun$)\dotfill                         & $0.770_{-0.028}^{+0.024}$ \\ [+2pt]
$R_{\star}$ ($\rsun$)\dotfill                         & $0.726_{-0.015}^{+0.018}$ \\ [+2pt]
$\log_{10}[L_{\star}/\lsun]$\dotfill                        & $-0.570_{-0.034}^{+0.036}$ \\ [+1pt]
$M_V$ (mag)\dotfill                                      & $6.53 \pm 0.12$ \\ [+1pt]
$M_{K_s}$ (mag)\dotfill                                  & $4.21 \pm 0.06$ \\ [+1pt]
Distance (pc)\dotfill                                    & $330 \pm 10$ \\ [+2pt]
Age (Gyr)\dotfill                                        & $3.3_{-0.8}^{+5.8}$
\enddata
\tablenotetext{a}{Value from SPC.}
\tablenotetext{b}{Mean stellar density constraint from transit light
  curve fits to \koib, \koic, and \koid\ (see text).}
\end{deluxetable}

Given these properties for \koi, we investigated whether the measured
brightness and color of the 2\farcs2 neighbor reported earlier are
consistent with those expected for a physically associated
main-sequence star of later spectral type, i.e., one falling on the
same Dartmouth isochrone as the primary. We find that an M4 or M5
dwarf with a mass around 0.20--0.21~$\msun$ would have
approximately the right brightness compared to the primary, though its
$J-K$ color would be about 0.16~mag bluer than we measure. However,
given the uncertainties that may be expected in the theoretical flux
predictions for cool stars (based here on PHOENIX model atmospheres
implemented in the Dartmouth models), as well as variations in color
that may occur in real stars due, e.g., to chromospheric activity, we
consider the measured properties to be still consistent with a bound
companion, although a chance alignment cannot be ruled out.

\section{STATISTICAL VALIDATION}
\label{sec:validation}

%% VALIDATION
%%

Transiting planet candidates require extra care to show that the
periodic dips in stellar brightness are not astrophysical false
positives, caused by other phenomena such as an eclipsing binary
blended with the target (a ``blend'').  Because \koi\ is a faint star
($V \approx 14.3$), it is challenging to confirm the planetary nature
any of the candidates in this system dynamically, by measuring the
Doppler shifts they induce on the host star. The alternative is to
validate them statistically, showing that the likelihood of a true
planet is far greater than that of a false positive. \cite{rowe:2014}
followed this approach and reported the validation of two of the
candidates, \koib\ and \koic, based on the argument that most
candidates in multiple systems can be shown statistically to have a
very high chance of being true planets \citep{lissauer:2012,
lissauer:2014}.  These two planets received the official designations
\kepb\ and \kepc. The validations relied in part on an
examination of existing follow-up observations including spectroscopy
and high-resolution imaging, and on an analysis of the flux
centroids. The other two candidates in the system, \koie\ and
\koid, were not considered validated by \cite{rowe:2014} because the
centroid information available was insufficient to determine whether
the source of the photometric signals coincided with the location of
the target, within errors.  Additionally, the period of \koie\ was
not precisely known, since only one transit had
occurred in the data at their disposal (Q1--Q10).

After the publication of the \cite{rowe:2014} work, AO imaging of
\koi\ was obtained that showed the presence of a 2\farcs2 companion
that was unknown at the time (see \S~\ref{sub:imaging}), and
could possibly be the source of one of the signals. However, as
pointed out earlier, the refined centroid information now available
that includes \kepler\ observations from Q1--Q17 firmly rules out that
the companion is causing the transits in \koib\ and \koic, as
it is well beyond the 3\,$\sigma$ exclusion regions for these candidates
(0\farcs396 and 0\farcs603, respectively;
\S~\ref{sub:centroids}). Thus, the validations of \cite{rowe:2014}
stand.

We describe below our efforts to validate the other two candidates,
\koie\ (the snow-line candidate) and \koid, using the
\blender\ technique \citep{torres:2004,torres:2011,fressin:2012}. This
procedure has been applied successfully to the validation of many
other transit candidates from \kepler\ \citep[for recent examples see,
e.g.,][]{meibom:2013,ballard:2013,kipping:2014,torres:2015,
jenkins:2015}. \blender\ addresses the possibility that the signals 
originate in an unseen background/foreground eclipsing binary (BEB) 
along the line of sight, a background or foreground star transited by 
a larger planet (BP scenario), or a stellar companion physically 
associated with the target that is in turn transited by another star or 
by a planet. These types of blends are usually the most difficult to 
rule out. The companions in the last two cases are usually close enough 
to the target as to be spatially unresolved. We refer to those 
hierarchical triple configurations as HTS or HTP, respectively, depending 
on the nature of the eclipsing object (star or planet).  Other types of 
false positives that do not involve contamination by another object along
the line of sight include grazing eclipsing binaries, and transits of
a small star in front of a giant star. However, these cases can be
easily ruled out as their signals would be inconsistent with the
observed durations of transit ingress and egress for the two
candidates.

Our validations with \blender\ follow closely the procedure described
by \cite{kipping:2014} or \cite{torres:2015}; the reader is referred to
these sources for details of the methodology. In essence,
\blender\ uses the shape of a transit light curve to rule out blend
scenarios that would lead to the wrong shape for a transit.  Large
numbers of false positives of different kinds are simulated, and the
synthetic light curves are then compared with the
\kepler\ observations in a $\chi^2$ sense. Blends giving poor fits to
the real data are considered to be excluded, and the ensemble of
results places tight constraints on the detailed properties of viable
blends including the sizes or masses of the objects involved, their
brightness and colors, the linear distance between the
background/foreground eclipsing pair and the KOI, and even the
eccentricities of the orbits.

\subsection{\koie, a Snow-line Candidate}
\label{sub:KOI-490e}

\begin{figure*}
\centering
\begin{tabular}{ccc}
\includegraphics[width=5.55cm]{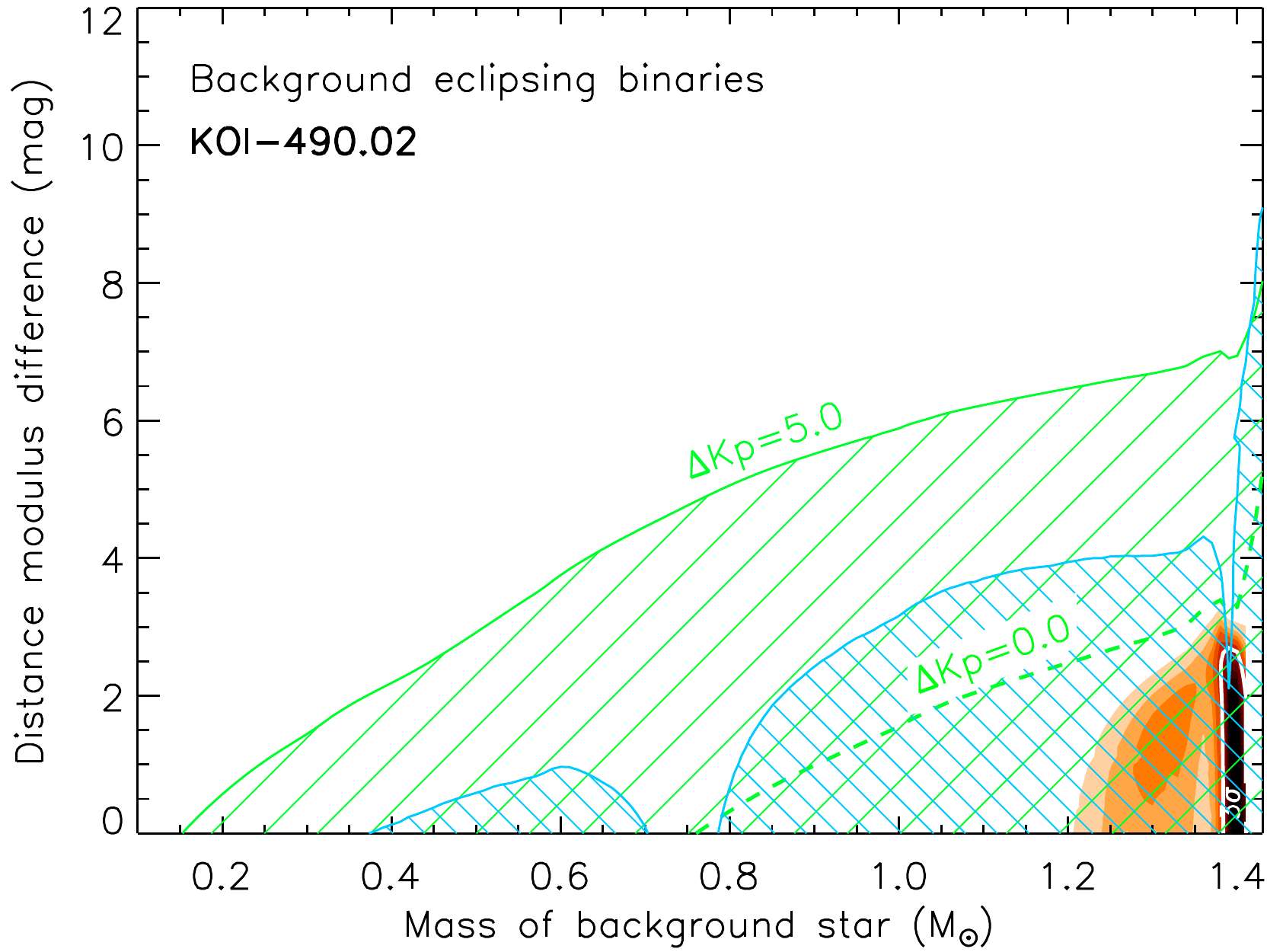} &
\includegraphics[width=5.55cm]{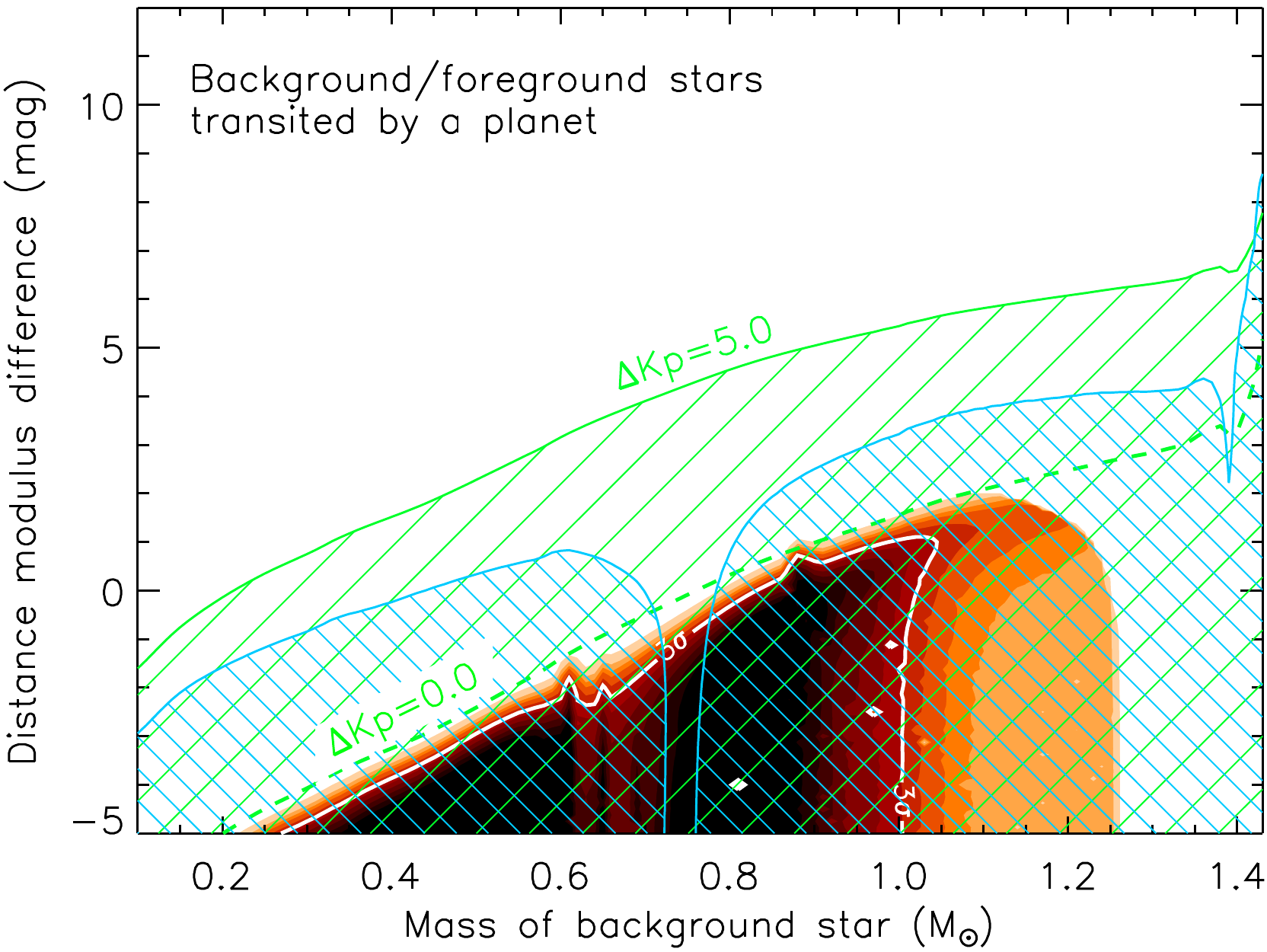} &
\includegraphics[width=5.55cm]{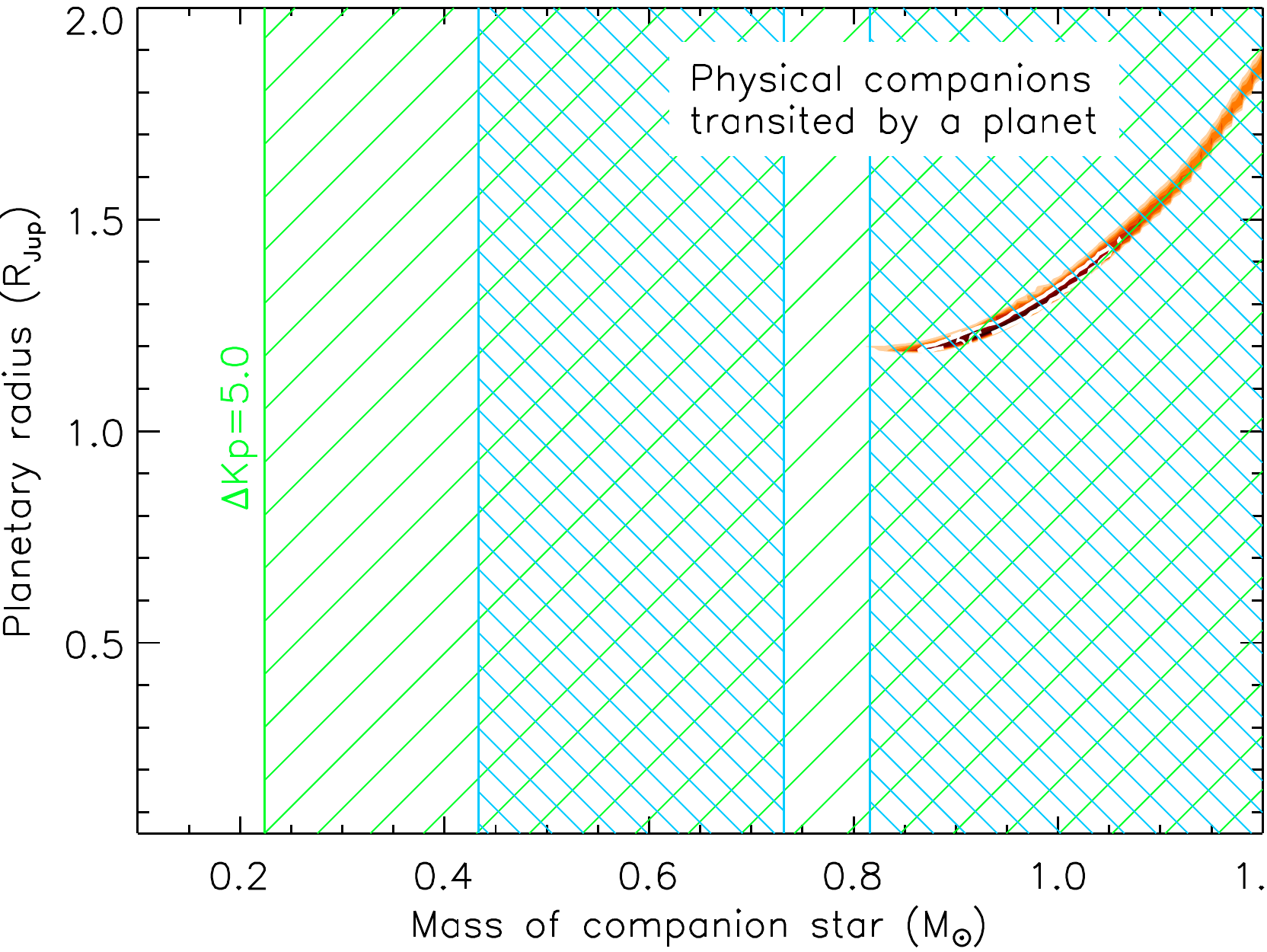} \\[1ex]
\end{tabular}
\figcaption[]{
Map of the $\chi^2$ surface (goodness of fit) for \koie\ for three 
different blend scenarios, as labeled. Only blends within the solid 
white contours (darker shading) provide fits to the \kepler\ light 
curves that are within acceptable limits \citep[3$\sigma$, where 
$\sigma$ is the significance level of the $\chi^2$ difference compared 
to a transiting planet model fit; see][]{fressin:2012}. Other 
concentric colored areas (lighter colors) represent fits that are 
increasingly worse (4\,$\sigma$, 5\,$\sigma$, etc.), which we consider
to be ruled out. The hatched green areas indicate regions of parameter 
space where blended stars can be excluded if they are within 0\farcs43 
of the target (half-width of the spectrometer slit), within five 
magnitudes in brightness (1\% relative flux), and have a radial velocity
differing from the target by 10~\kms\ or more. In all of the above cases
they would have been detected spectroscopically.  Blends in the hatched
blue areas can also be ruled out because they would be either too red 
(left) or too blue (right) compared to the measured $r-K_s$ color of 
\koi, by more than three times the measurement uncertainty. 
\emph{Left:} BEB scenario. The vertical axis represents the linear 
distance between the eclipsing binary and the target ($D_{\rm BEB} - 
D_{\rm targ}$), cast for convenience in terms of the distance modulus 
difference $\Delta\delta = 5 \log(D_{\rm BEB}/D_{\rm targ})$. The dashed 
green line shown for reference is the locus of blends of equal apparent 
$K\!p$ brightness as the target.
\emph{Middle:} BP scenarios. As before, only blends that are brighter 
than the target (below the dashed green line) are able to mimic the 
light curve. The $r-K_s$ color constraint rules out most of those. 
\emph{Right:} HTP scenario. The vertical axis now shows the size of the 
planet transiting the companion star, in units of Jupiter's radius. All 
blends of this kind that provide acceptable fits to the light curve are 
too blue, and are therefore ruled out.
\label{fig:blender.koi490e}}
\end{figure*}

Our simulations with \blender\ rule out background eclipsing binaries
as the source of the signal. This is illustrated in
Figure~\ref{fig:blender.koi490e} (left panel), where we show the
$\chi^2$ landscape in a representative cross-section of parameter
space. The diagram shows the linear separation between the BEB and the
target as a function of the mass of the primary star in the BEB. The
only scenarios of this kind that provide acceptable fits to the
\kepler\ light curve are those in which the main star of the binary is
about twice as massive as \koi\ (i.e., $M \sim 1.4~\msun$). These
false positives occupy a narrow vertical strip on the lower right
corner of the first panel in the figure (darker region contained
within the white, 3$\sigma$ contour).  However, all of these
configurations result in a combined $r-K_s$ color index for the blend
that is much bluer than the measured value for \koi\ ($r-K_s = 2.095
\pm 0.027$)\footnote{This accounts for zero-point corrections to the
Sloan magnitudes in the KIC, as prescribed by 
\cite{pinsonneault:2012}.}, as indicated by the hatched blue region in
the figure within which all blends have the wrong color. Furthermore,
in these false positive configurations with F-type primaries the BEB
is brighter than the target itself (see dashed green line). This
conflicts with our spectroscopic classification of \koi\ as an early K
dwarf. We conclude that BEBs cannot mimic the transits of \koie\ and
simultaneously satisfy all observational constraints. This also rules
out the 2\farcs2 companion as the source of the transits.

Blends involving background or foreground stars transited by a larger
planet (BP scenario) are more easily able to match the transit shape
and depth. This is shown in the middle panel of
Figure~\ref{fig:blender.koi490e}, in which the permitted region is
larger and accommodates chance alignments with stars between about
0.25~$\msun$ and 1.0~$\msun$. The spectroscopic constraint
represented by the hatched green area excludes all such blends if the
intruding stars have $\Delta K\!p < 5$~mag and fall within the
spectrometer slit (unless their spectral lines are blended with those
of the target), as we would have detected them in our Keck/HIRES
observation. Most other scenarios are also ruled out by the color
constraint, but there is a narrow strip of viable blends in which the
contaminating star has the same color (mass) as the target (see
figure) so that it does not alter the combined $r-K_s$ index. These
would then be near twin stars of our target, and they would have to be
brighter than our nominal target because they are in the
foreground. However, a star so similar to our target that is transited
by a planet and is along the same line of sight but is brighter would
effectively be our target, so we do not consider this as a false
positive.

\blender\ indicates that physical companions eclipsed by another star
(HTS scenario) invariably have the wrong shape for the transit, or
produce secondary eclipses that are not seen in the \kepler\ data for
\koie. Even in cases that show only a single eclipse due to a high
eccentricity and a special orientation \citep{santerne:2013} the
properties of the primary of the eclipsing binary would be such that
the overall brightness would make the binary detectable and/or make
its color inconsistent with the measured color index of the target.
These configurations are therefore easily ruled out. Physically
associated stars transited by a larger planet (HTP scenario) can mimic
the light curve only for a very narrow range of parameters, as
illustrated in Figure~\ref{fig:blender.koi490e}, but those blends
are all too blue because the companion needs to be even more massive
than the target in order to produce the right shape for the transit,
after accounting for dilution.  We can thus exclude this category of
blends completely.

In summary, our \blender\ simulations for \koie\ combined with the
observational constraints allow us to easily validate the candidate as
a bona fide planet, ruling out as the source of the transits not only
unseen background stars but also the known 2\farcs2 companion. A
significant factor aiding in this process is the very high
signal-to-noise ratio of the deep transits, thanks to which the shape
is so well defined (particularly the ingress/egress phases) that very few
configurations involving another star along the line of sight can
match the \kepler\ photometry as well as a true transiting planet fit.

\subsection{\koid}
\label{sub:KOI-490d}

\begin{figure*}
\centering
\begin{tabular}{ccc}
\includegraphics[width=5.55cm]{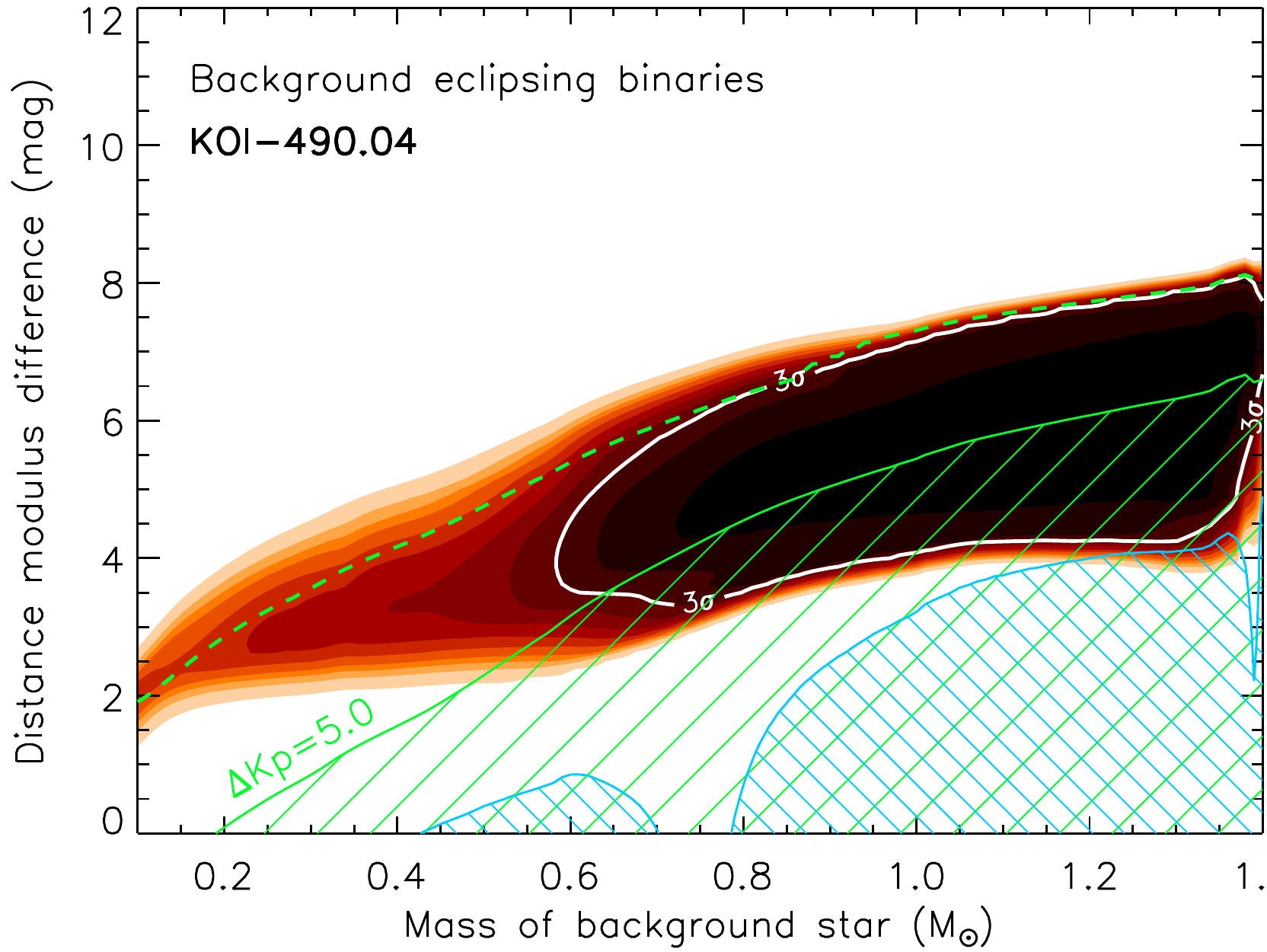} &
\includegraphics[width=5.55cm]{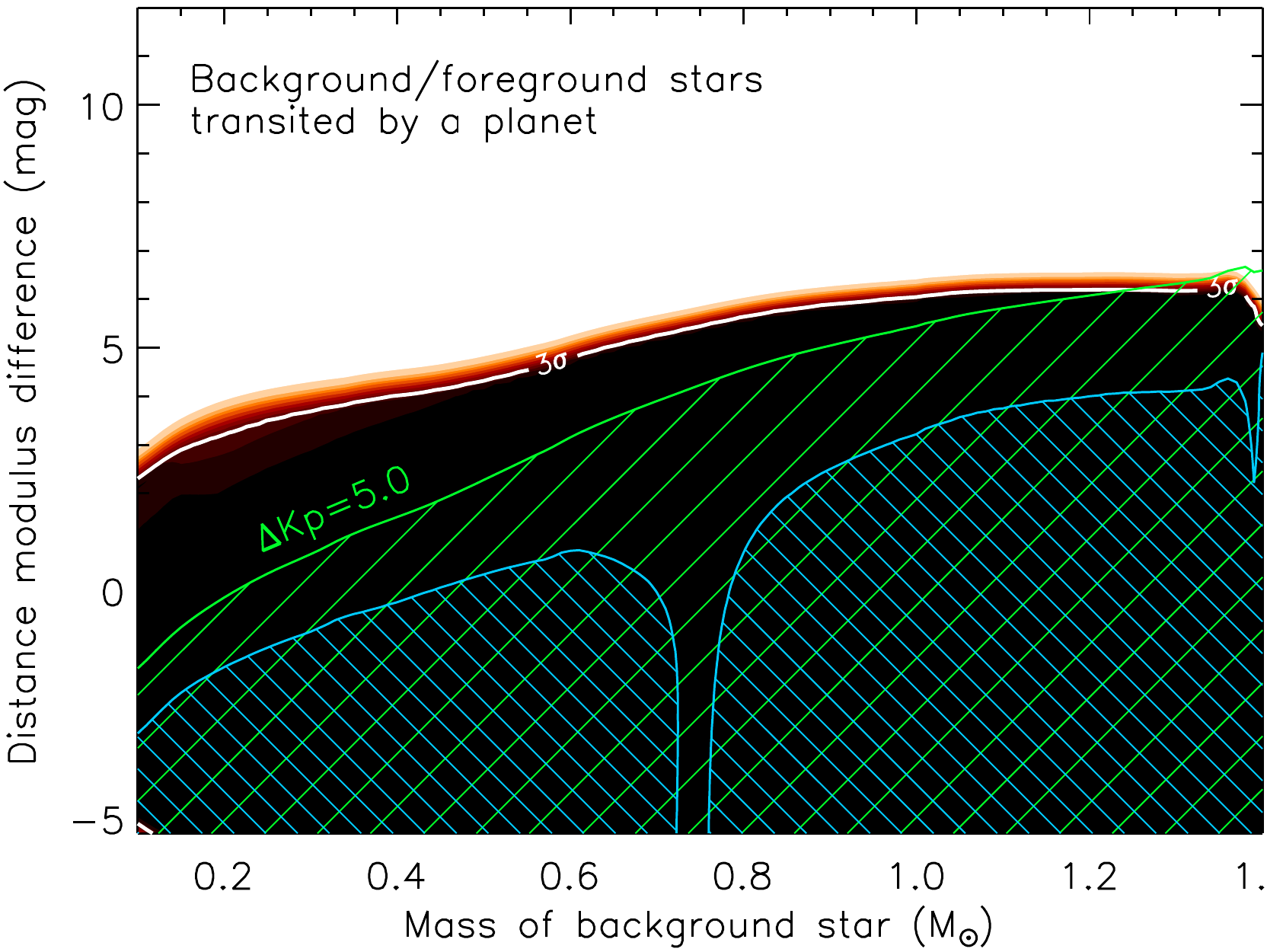} &
\includegraphics[width=5.55cm]{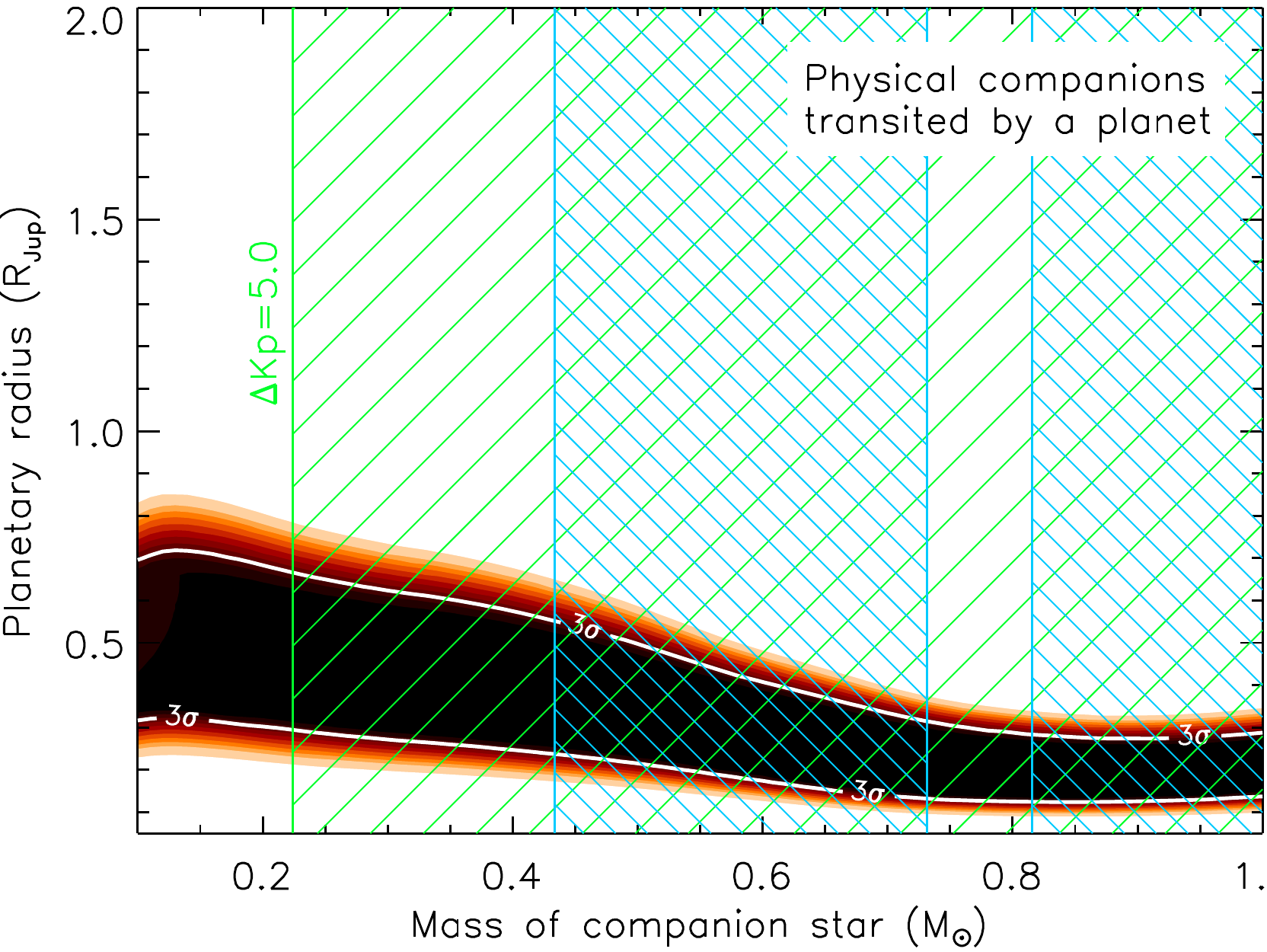} \\[1ex]
\end{tabular}
\figcaption[]{
Similar to Figure~\ref{fig:blender.koi490e} for \koid, using the 
same color scheme. For this candidate all three scenarios feature blends 
that cannot be ruled out from the shape of the transits or constraints 
from our follow-up observations (darker areas not overlapping the hatched
regions). The expected frequencies of each of these types of blends are 
estimated in the text.
\label{fig:blender.koi490d}}
\end{figure*}

The transits of this candidate are much shallower than those of
\koie, and as a result the constraint on the detailed shape of the
light curve provided by the \kepler\ data is considerably weaker. Our
\blender\ simulations indicate that false positives involving a bound
companion eclipsed by a smaller star (HTS scenario) do not provide
acceptable fits to the light curve, as in the previous case. However,
not all blends corresponding to the BEB, BP, and HTP configurations
can be ruled out. We illustrate this in
Figure~\ref{fig:blender.koi490d}. For example, background eclipsing
binaries that are more than five magnitudes fainter than the target in
the \kepler\ band can match the shape of the light curve just as well
as a model of a planet transiting the target, for a wide range of
masses of the primary star of the binary between 0.6~$\msun$ and
1.4~$\msun$ (left panel). An even wider range of masses is
permitted for background stars transited by a larger planet (BP,
middle panel). Similarly, small stars physically bound to the target
can mimic the light curve closely if transited by a planet of suitable
size (right panel, HTP). Our follow-up observations may rule out some
fraction of these blends, but not all of them.

To compute the expected rates of occurrence of each of these
scenarios we followed the Monte Carlo procedure described by
\cite{torres:2015}, in which we simulated large numbers of blends and
counted those that satisfy the constraints from \blender\ and the
follow-up observations.  The technique relies on the number densities
of stars in the vicinity of \koi\ (at Galactic latitude +9\fdg4), the
estimated rates of occurrence of eclipsing binaries and transiting planets
of various sizes and orbital periods, and other known properties of
binary stars such as the distributions of their periods,
eccentricities, mass ratios, etc.  Details of these calculations may
be found in the work cited above. We obtained estimated frequencies of
$4.17 \times 10^{-8}$ for the BEB scenario, $1.09 \times 10^{-7}$ for
BP, and $4.10 \times 10^{-6}$ for HTP configurations. The total blend
frequency is the sum of these, or $4.25 \times 10^{-6}$. While this
value may seem small in absolute terms, the a priori rate of
occurrence of transiting planets of a given period and size (``planet
prior'', PL) is also expected to be small. For a secure validation we
require here that the ``odds ratio'' ${\rm
  PL}/(\rm{BEB}+\rm{BP}+\rm{HTP})$ be large enough so that the planet
hypothesis is clearly favored over a false positive. Our estimate of
the planet prior is $2.02 \times 10^{-3}$, based on the number of
known KOIs of similar size and period as the candidate. The resulting
odds ratio is then 475, which corresponds to a confidence level of
99.79\% that the signal is due to a bona fide planet. As this exceeds
the 3$\sigma$ significance threshold typically adopted in
\blender\ applications, it formally validates \koid\ as a planet.
There is, however, an important caveat to make: an implicit assumption
for \blender\ is that there is no visible sign of a blend, whereas we
know of the presence of the 2\farcs2 companion. Our
\blender\ simulations in fact show
(Figure~\ref{fig:blender.koi490d}) that a faint star such as this
could well be causing the signal if transited by a larger planet, both
as a physical companion to \koi\ or as a background/foreground
interloper. Under these circumstances the validation with \blender\ is
not sufficient as it applies only to unseen sources, and we must seek
alternate ways of ruling out the 2\farcs2 companion as the cause of
the transit signals. This is successfully achieved through the use
of asterodensity profiling, as described in \S\ref{sub:fits_d}.

\section{LIGHT CURVE FITS}
\label{sec:fits}

%% FITS
%%

\subsection{Joint fit to \kepb\ and \kepc}
\label{sub:fit_bc}

Planets \kepb\ and \kepc\ were both validated by \citet{rowe:2014} and new
centroid information since that time excludes the possibility of these
two objects orbiting the 2\farcs2 companion. This therefore establishes
that \kepb\ and \kepc\ orbit the same star, namely the target star \kep,
to high confidence.

Fitting the transit light curve of a planet includes multiple parameters
pertaining to the star itself, which may be described by the vector
$\thetastar$. The vector $\thetastar$ contains the limb darkening coefficients
describing the stellar intensity profile and the mean stellar density,
$\rho_{\star}$. Since \kepb\ and \kepc\ orbit the same star, we can fit the
transit lightcurves of both simultaneously, adopting a global $\thetastar$.
By conditioning $\thetastar$ on the data describing both planets, we obtain a
higher signal-to-noise measurement of these terms, which in turn leads to
somewhat better precision on the local parameters, $\thetaplan$, describing
each planet (due to the inter-parameter covariances \citealt{carter:2008}).

In this work, we use the quadratic limb darkening law with the optimal
parameterization ($q_1$ \& $q_2$) described by \citet{LDfitting:2013}. The light
curves are generated with the \citet{mandel:2002} algorithm using 30-point
resampling to account for those point which are long-cadence, as described in
\citet{binning:2010}. Quarter-to-quarter contamination factors are accounted for,
using the ``CROWDSAP'' header information in the raw fits files via the method
described by \citet{nightside:2009}. A blending factor affecting all quarters
due to the 2\farcs2 companion is also accounted for via this method.

For the global blending factor, we convert the $J$ and $K_s$ colors observed in
the NIRC2 AO images to a \kepler\ bandpass magnitude using the fifth-order polynomial
relation in Appendix~A of \citet{howell:2012}. Assuming either a dwarf or a giant
leads to the same result (within the estimated uncertainty) of $K\!p = 19.0\pm0.1$.
Assuming Gaussian errors on the $J$ and $K_s$ colors from AO and adding in
quadrature an extra Gaussian uncertainty reflecting the 0.05 magnitude error in the
\citet{howell:2012} relation, we estimate a blending factor of
$\log\beta = (-1.976\pm0.036)$, where $\beta$ is the flux ratio between the target
and the companion in the \kepler\ bandpass. This is treated as a normal prior in our
fits and added to the $\thetastar$ vector, since it is a term affecting all of the
planets. The parameters and priors used are listed in Table~\ref{tab:priors},
with the $\thetastar$ plus two sets of $\thetaplan$ parameters giving a total
of 12 free parameters in our model.

\begin{table}
\caption{
Priors for the joint fit to \koib\ \& \koic. $\mathcal{J}$ is a Jeffreys
prior, $\mathcal{N}$ is a normal and $\mathcal{U}$ is uniform.
}
\centering
\begin{tabular}{l c c}
\hline\hline
Description & Symbol & Prior \\ [0.5ex]
\hline
\it{Global parameters} & $\thetastar$ \\
\hline
Mean stellar density [kg\,m$^{-3}$] & $\rho_{\star}$ & $\mathcal{J}[10,10^6]$ \\
Limb darkening coefficient 1 & $q_1$ & $\mathcal{U}[0,1]$ \\
Limb darkening coefficient 2 & $q_2$ & $\mathcal{U}[0,1]$ \\
Log of the blending flux ratio & $\log_{10}\beta$ & $\mathcal{N}(-1.976,0.036)$ \\
\hline
\it{Local parameters} & $\thetaplan$ \\
\hline
Ratio-of-radii & $(R_P/R_{\star})$ & $\mathcal{U}[0,1]$ \\
Impact parameter & $b$ & $\mathcal{U}[0,2]$ \\
Orbital period [d] & $P$ & $\mathcal{U}[\bar{P}-0.1,\bar{P}+0.1]$ \\
Time of transit minimum [d] & $\tau$ & $\mathcal{U}[\bar{\tau}-0.1,\bar{\tau}+0.1]$ \\ [1ex]
\hline
\end{tabular}
\label{tab:priors}
\end{table}

Fits were achieved using the multimodal nested sampling algorithm \multi\
\citep{feroz:2008,feroz:2009} with 4000 live points and a target efficiency set
to 0.1. The eccentricities of \kepb\ and \kepc\ are assumed to be zero in the
fits. Multi-planet \kepler\ systems are known to have low eccentricities, with
\citet{vaneylen:2015} finding a Rayleigh distribution with $\sigma_e=0.049\pm0.013$
describes the overall population. Moreover, \kepb\ and \kepc\ orbit the same star
with orbital periods of 4.4\,d and 7.4\,d, placing them in close proximity to both
the star and each other. We therefore expect these planets to be particularly
likely to have near-zero eccentricity, from a dynamical perspective.

\begin{figure}
\begin{center}
\includegraphics[width=8.4 cm]{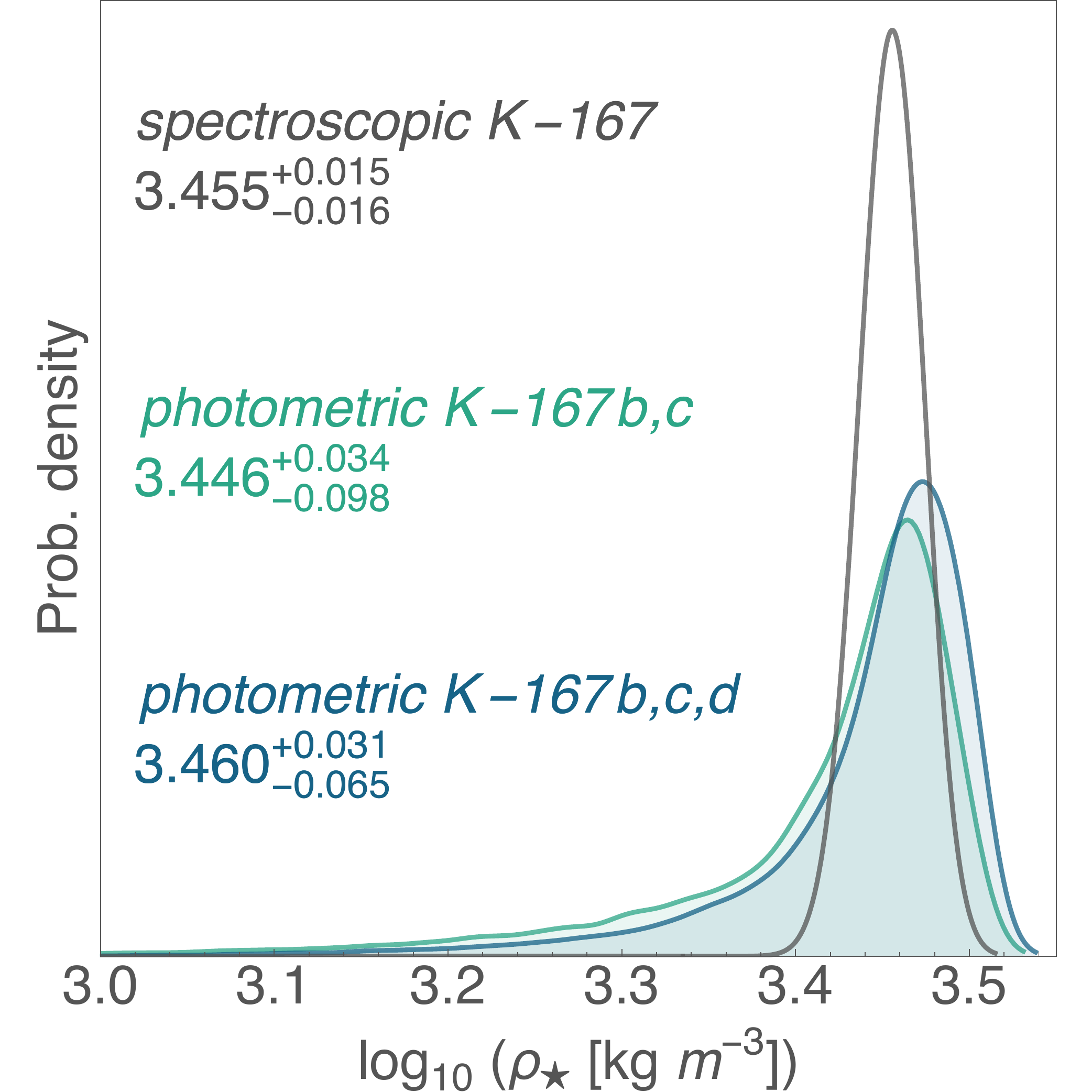}
\caption{
Posterior distribution of the mean stellar density of \kep\ conditioned on the
transits of \kepb\ \& c (green) and \kepb, c \& d (blue). We only
assume that the planets orbit the same star and have circular orbits. For
comparison, the posterior derived using isochrone matching of the SPC
stellar atmosphere constraints is shown in black.
}
\label{fig:logrhostar}
\end{center}
\end{figure}

Rather than describe the posteriors found for each parameter, we focus here on
the term $\rho_{\star}$, since the others will be superseded by the global fits
performed later. The posterior distribution for $\log_{10}(\rho_{\star})$ (it is more
appropriate to discuss the log since we invoked a Jeffreys prior) yields
$\log_{10}[\rho_{\star} ($kg\,m$^{-3})] = 3.446_{-0.098}^{+0.034}$ and is plotted in
Figure~\ref{fig:logrhostar}. This may be compared to the density expected by
iscohrone-matching using the effective temperature, metallicity and surface gravity
found using SPC. Drawing random samples from three normal distributions
describing each and finding the nearest Dartmouth isochrone each time, we derive a
wholly independent stellar density for \kep\ of
$\log_{10}[\rho_{\star} ($kg\,m$^{-3})] = 3.455_{-0.016}^{+0.015}$. The close agreement between
the two is further evidence that \kepb\ and \kepc\ orbit the target star, although
their validation \citep{rowe:2014} was performed independent of this fact.

\subsection{Two Fits for \koid}
\label{sub:fits_d}

From the validation discussion in \S~\ref{sec:validation}, it was established that
\koid\ does not orbit an unseen companion to 3\,$\sigma$ confidence, but may still
orbit the seen 2\farcs2 companion. Here, we perform two fits to explore these
two hypotheses.

In hypothesis A ($\mathcal{H}_A$), we assume that \koid\ orbits the target star, which
we have established also hosts \kepb\ and \kepc. The posterior distribution of the
light curve derived stellar density from the fit in \S~\ref{sub:fit_bc} becomes the
prior for the same term in this hypothesis. Note that this distribution does not
invoke any information from the spectroscopic analysis. All of the other parameters retain
the same priors listed in Table~\ref{tab:priors}. Specifically, the limb darkening
parameters are treated as free again, since these were poorly constrained from the
previous fit.

For $\mathcal{H}_A$, we relax the assumption of a circular orbit.
Since \koid\ is further from both the star and the other two planets 
($P\sim21.8$\,d), higher eccentricities are possible. In this hypothesis though,
\koid\ belongs to a typical \kepler\ multi-planet system and thus we adopt the eccentricity
distribution derived by \citet{vaneylen:2015}; a Rayleigh distribution with
$\sigma_e=0.049$. The prior for the argument of periapsis becomes increasingly
non-uniform as eccentricity diverges from zero, due to a geometric effect (see
\citealt{eprior:2014}). Nevertheless, a uniform prior is reasonable in this case given
the low-eccentricity nature of the \citet{vaneylen:2015} distribution.

In hypothesis B ($\mathcal{H}_B$), we assume that \koid\ orbits the 2\farcs2 companion.
Here, the object can no longer be considered to belong to a multi-planet system, since
\kepb\ and \kepc\ orbit a different star now. Therefore, the potential for high
eccentricities becomes even greater and we consider the object to follow a Beta
distribution calibrated to the population of radial velocity planets with orbital
periods less than one year, as described by \citet{beta:2013} ($\sim 
\mathrm{Beta}[0.694,3.252]$). In this case, the geometric bias due to the transit
probability is more significant and requires accounting for. We therefore use the
\eccsamples\ code described by \citet{eprior:2014} to modify the Beta prior to a prior
conditioned on the fact we know this object is transiting. However, rather than
directly sampling from the prior, we penalized the likelihood function appropriately
to improve computational efficiency. Additionally, since the star is different to \kep,
the mean stellar density follows an uninformative Jeffreys prior,
$\mathcal{J}[1,10^9]$\,kg\,m$^{-3}$. Finally, the blending prior is flipped to consider
the blend source being \kep.

We fitted both models using \multi, which returns the Bayesian evidence, 
$\mathcal{Z}$, enabling Bayesian model selection between the two. Note that this 
is essentially a more advanced treatment of using the photo-blend effect 
\citep{AP:2014} employed in validating several candidates by \citet{torres:2015}. 
Bayesian model selection favors hypothesis A with $\Delta(\log\mathcal{Z}) = 
(5.57\pm0.10)$. Given that only two hypothesis exist, the statistical 
significance of hypothesis A being the preferred model is 2.9\,$\sigma$. 
We therefore use the photo-blend effect to show that \koid\ orbits
the target star to 99.6\% confidence.

However, in hypothesis B, the mean stellar density required to explain the
data is $\log_{10}[\rho_{\star} ($kg\,m$^{-3})] = 3.49_{-0.28}^{+0.16}$.
This would make the companion star of similar spectral type to that of \kep.
This essentially excludes the possibility of a bound binary, since the AO
companion could not be 5 magnitudes fainter in this case. If this were
a chance alignment of a background star, we would still expect the host star
to have a similar color to the primary. However, $J-K_s$ of \kep\ is
$(0.60\pm0.03)$ whereas the AO companion has $J-K_s = (0.94\pm0.05)$.
Reddening can not explain such a large difference either and thus we
conclude that hypothesis B is even less likely than indicated by
the formal Bayesian evidence comparison. We therefore conclude that
\koid\ orbits \kep\ to a confidence exceeding 3\,$\sigma$ and thus may
be considered a ``validated'' planet, designated \kepd.

\begin{figure}
\begin{center}
\includegraphics[width=8.4 cm]{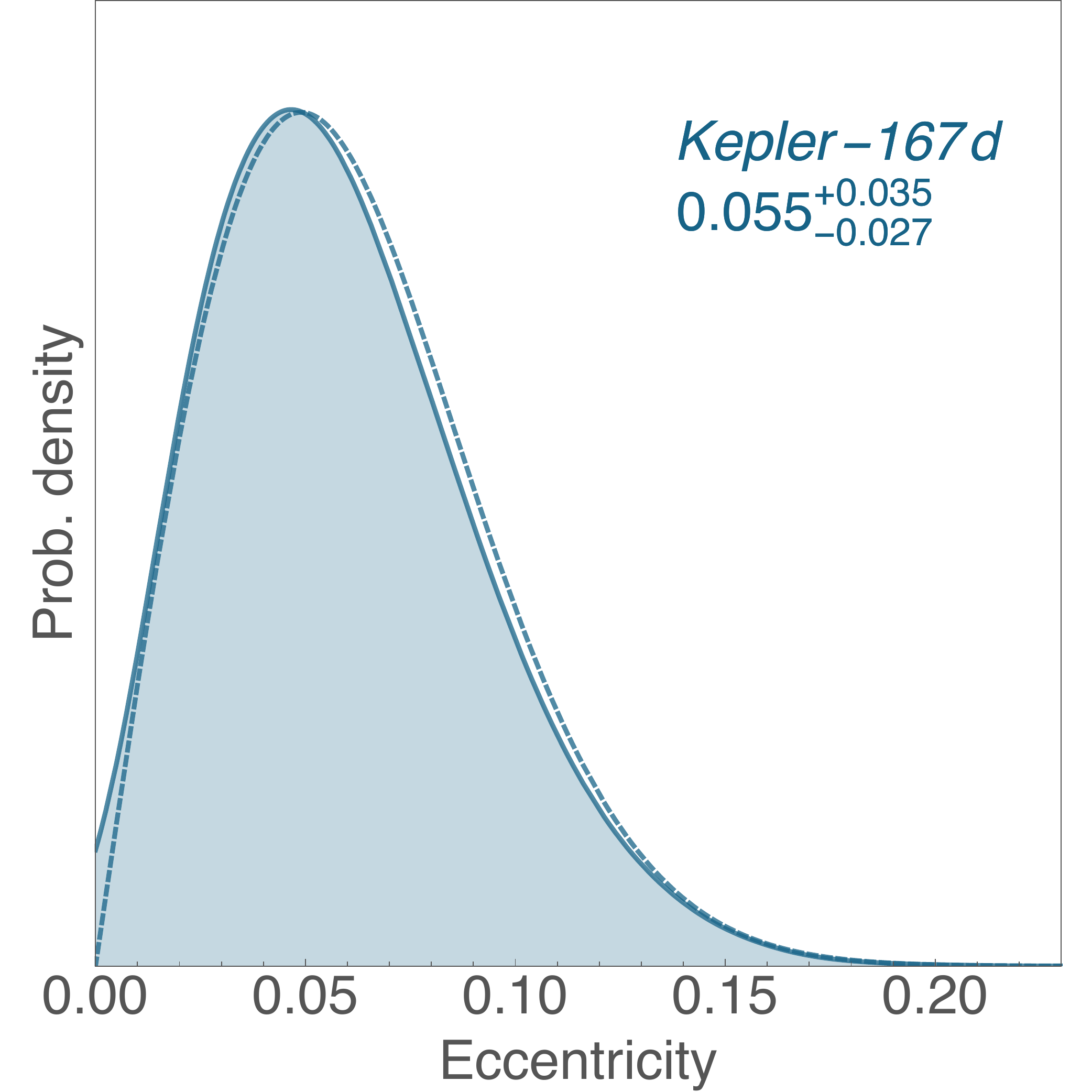}
\caption{
Posterior distribution (solid) of the orbital eccentricity of \kepd,
with comparison to the prior (dashed) describing \kepler\ multis
\citep{vaneylen:2015}. The eccentricity is constrained by the
comparison of the transit light curve shape of \kepd\ to the $\rho_{\star}$
constraint from the earlier joint fit of \kepb\ \& c.
}
\label{fig:eccentricity_d}
\end{center}
\end{figure}

Since hypothesis A invokes the posterior of $\rho_{\star}$ from the
earlier joint fit of \kepb\ \& c as a prior, the eccentricity of
\kepd\ is constrained without any use of stellar evolution or atmosphere
models. This demonstrates perhaps the first applied example of
Multibody Asterodensity Profiling (MAP), proposed by \citet{MAP:2012}.
However, we find that the light curve of \kepd\ is of insufficient data quality
to provide a meaningful improvement on the eccentricity constraint
over the prior. Specifically, in hypothesis A, the eccentricity 
posterior closely resembles the prior (see Figure~\ref{fig:eccentricity_d})
and the credible intervals on $e$ change from $0.058_{-0.029}^{+0.036}$
(the prior) to $0.055_{-0.027}^{+0.035}$ (the posterior).

\subsection{Parameters and Eccentricity of \kepe}
\label{sub:fits_e}

\begin{figure*}
\begin{center}
\includegraphics[width=18.0 cm]{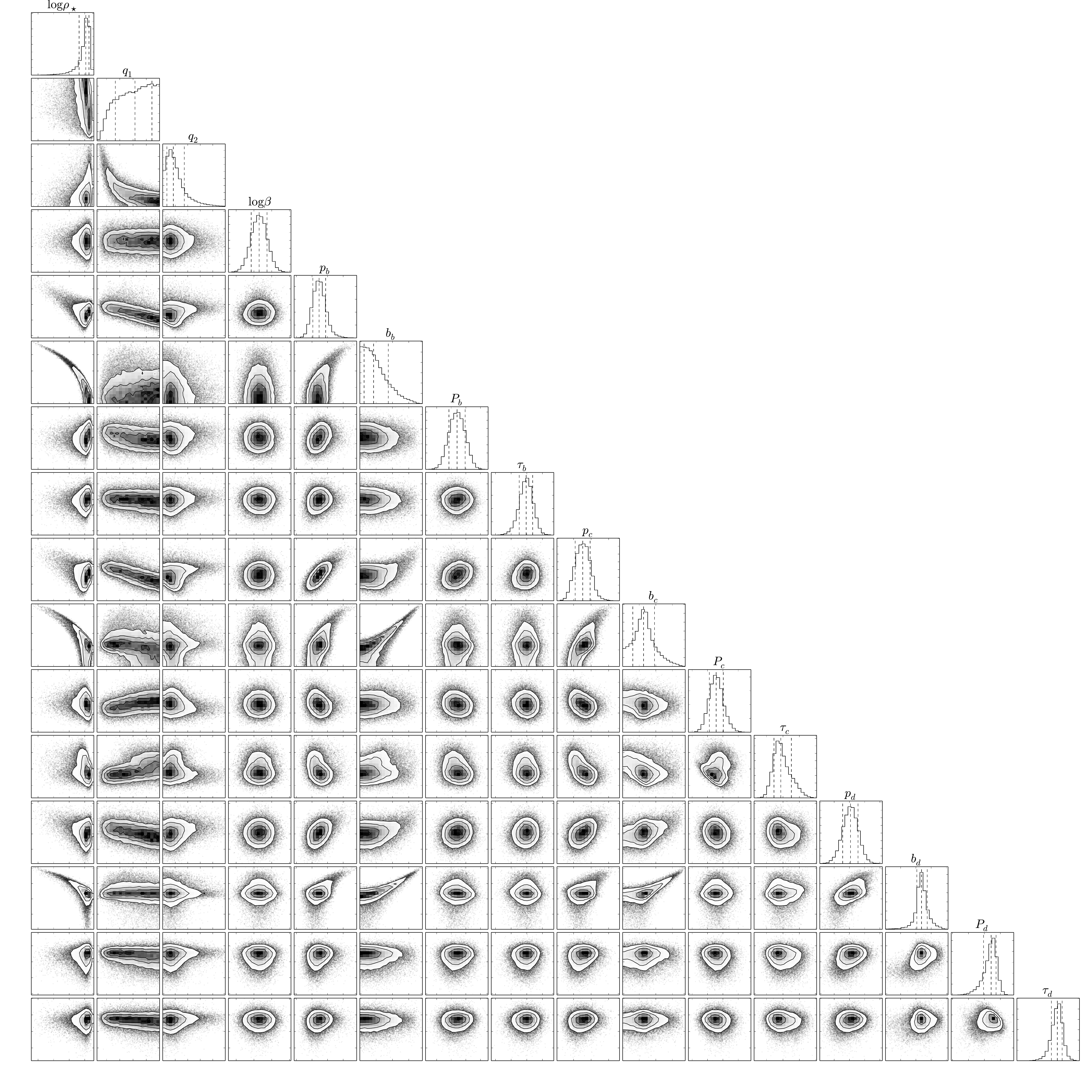}
\caption{
Triangle plot of the posterior distributions of the 16 parameters explored
in the joint fit of \kepb, c \& d. Contours mark the 0.5, 1.0, 1.5 \&
2.0\,$\sigma$ confidence intervals and dashed lines on the histograms mark
the median and surrounding 1\,$\sigma$ confidence interval. Posteriors may
be downloaded at \wwwcoolworlds.
}
\label{fig:K167bcd_post}
\end{center}
\end{figure*}

Having established that \kepb, \kepc\ and \kepd\ orbit the same star,
we perform a new global fit adopting a common $\thetastar$ for the
parent star. Since the eccentricity of \kepd\ is inferred to be
consistent with a low eccentricity prior, we assume the orbit is
nearly circular in this global fit. Including the third planet
provides a modest improvement in the constraint on the mean stellar
density, as expected. This may be seen in Figure~\ref{fig:logrhostar}
where the constraint tightens up to
$\log_{10}[\rho_{\star} ($kg\,m$^{-3})] = 3.460_{-0.065}^{+0.031}$.
We use this posterior on the mean density as an additional constraint
for the fundamental stellar parameters through isochrone matching,
as described earlier in \S~\ref{sub:stellar}. The full set of posteriors
from this fit are shown in Figure~\ref{fig:K167bcd_post} and are
available for download at \wwwcoolworlds.

\begin{figure*}
\begin{center}
\includegraphics[width=18.0 cm]{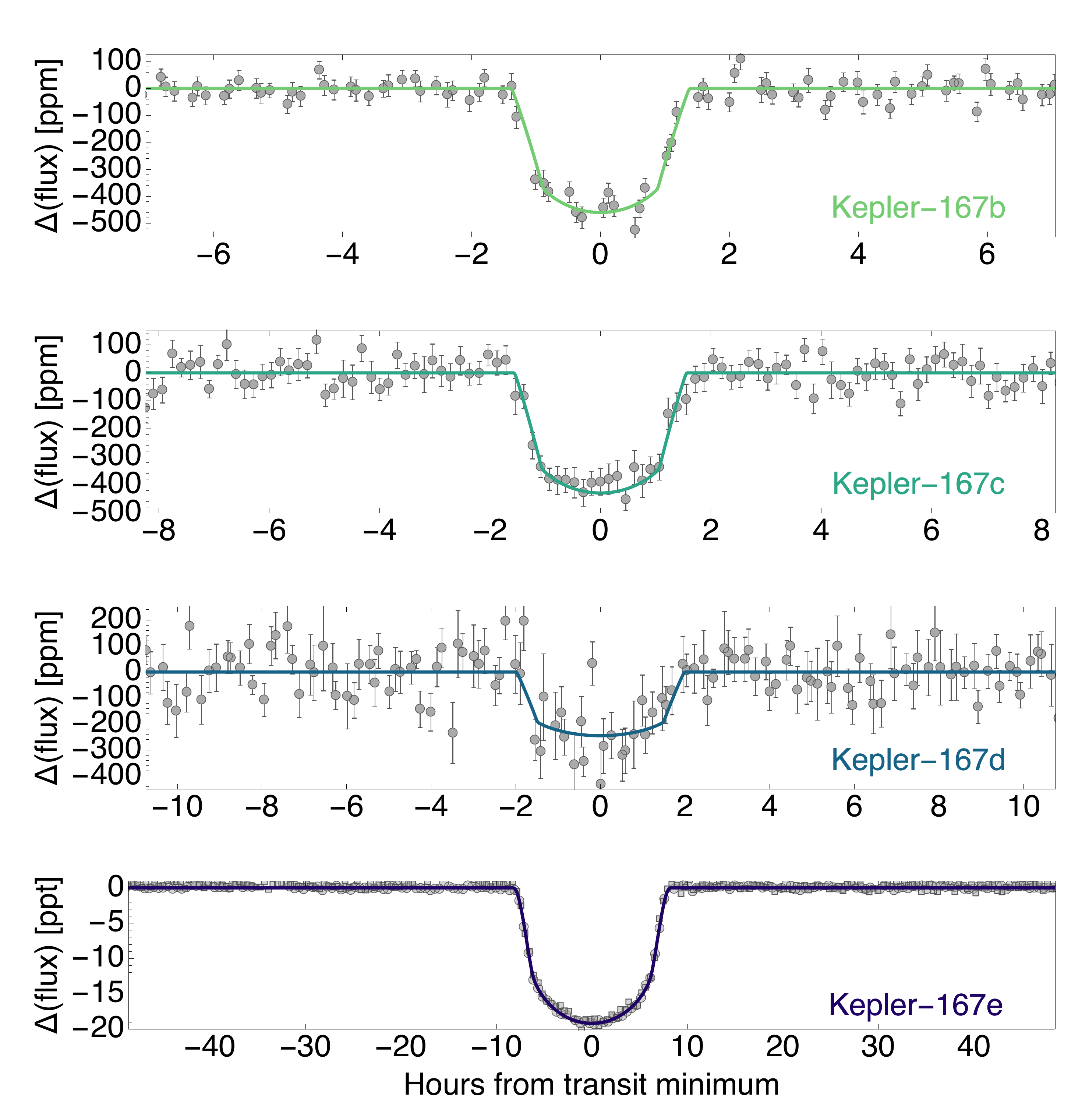}
\caption{
Folded transit light curves of \kepb, \kepc, \kepd\ and \kepe.
For the upper three, data (gray points) are binned to a 10\,minute cadence.
Light curve of \kepe\ uses 30\,minute binning and uses circles to denote
the first transit (Q4) and squares to denote the second transit (Q16). Note
that all of the transits were fitted using the original unbinned data.
}
\label{fig:lightcurves}
\end{center}
\end{figure*}

Using the new revised isochrone modeling and the ratio-of-radii posteriors
from this three-planet joint fit, we infer our best-estimate of the
radii of planets \kepb, c \& d to be 
$R_b = 1.615_{-0.043}^{+0.047}$\,$\rearth$,
$R_c = 1.548_{-0.048}^{+0.050}$\,$\rearth$ and
$R_d = 1.194_{-0.048}^{+0.049}$\,$\rearth$. The maximum likelihood
light curve models for the planets in the \kep\ system are shown in
Figure~\ref{fig:lightcurves}.

Note that despite using three transiting planets, the limb darkening
coefficients are poorly constrained returning only marginally different
posteriors from the priors. Specifically, we measure $q_1 = 
0.63_{-0.30}^{+0.26}$ and $q_2= 0.17_{-0.10}^{+0.18}$.

We now turn our attention to the outer planet, \kepe, which was 
validated earlier to orbit the target star (see \S~\ref{sub:KOI-490e}).
We first verified that the period of 1071\,d is the correct one, by inspecting
the raw light curve around integer ratios of the candidate period. Given
the depth of 1.6\%, even a simple visual inspection excludes this
possibility. As was done by \citet{kipping:2014}, we also verified that
the shape of the two transit events observed are consistent which is also
visually evident in Figure~\ref{fig:lightcurves}. Basic parameters derived
from two independent fits of each event reveal that the events are
consistent, as shown in Table~\ref{tab:indivparams} (note that
the first transit is long-cadence and the second, short).

\begin{table}
\caption{
Comparison of the basic transit parameters of \kepe\ when epochs 1 \&
2 are fitted independently. The last three differ by 0.9, 1.2 \& 1.2\,$\sigma$
respectively.
} % title of Table
\centering % used for centering table
\begin{tabular}{c c c} % centered columns (7 columns)
\hline\hline
Parameter & Epoch 1 & Epoch 2 \\ [0.5ex] % inserts table
%heading
\hline
$\tau$ [BKJD$_{\mathrm{UTC}}$-2,455,000] \dotfill & $253.28698_{-0.00043}^{+0.00042}$ & $1324.51928_{-0.00045}^{+0.00044}$ \\ %& - \\
$(R_P/R_{\star})$ \dotfill & $0.1265_{-0.0011}^{+0.0014}$ & $0.1284_{-0.0016}^{+0.0015}$ \\ % & 0.9\,$\sigma$ \\
$T_{14}$\,[hours] \dotfill & $16.241_{-0.079}^{+0.090}$ & $16.114_{-0.071}^{+0.070}$ \\%& 1.2\,$\sigma$ \\
$T_{23}$\,[hours] \dotfill & $12.48_{-0.18}^{+0.14}$ & $12.15_{-0.20}^{+0.20}$ \\%& 1.2\,$\sigma$\\ [1ex]
\hline %inserts single line
\end{tabular}
\label{tab:indivparams} % is used to refer this table in the text
\end{table}

Given that \kepe\ is a much longer orbital period planet than the
other three, the potential for an eccentric orbit is much higher both
a-priori as a member of the long-period planet sample \citep{beta:2013}
and dynamically since it is essentially decoupled from the other
three.

We treat the posterior distribution for the mean stellar density
of the \kepb, c \& d joint fit as a prior in the fits of planet e.
Although there is a weak constraint on the limb darkening coefficients,
we consider the information too weak to be worth including and thus
treat the limb darkening coefficients as independent and free. Therefore
the priors largely follow those listed in Table~\ref{tab:priors}, except
for $\rho_{\star}$.

It is also crucial to include the orbital eccentricity and argument of
periapsis. Given the potential for a large eccentricity, the geometric
bias effect described by \citet{eprior:2014} becomes pronounced and must
be accounted for. As was done in hypothesis B of the \koid\ fits, we use
the geometry-corrected Beta prior from \citet{eprior:2014} via a
likelihood penalization implementation, adopting Beta shape parameters
calibrated to the long-period ($>1$\,year) radial velocity exoplanet
catalog \citep{beta:2013} (specifically $\sim\mathrm{Beta}[1.12,3.09]$).

\begin{figure}
\begin{center}
\includegraphics[width=8.4 cm]{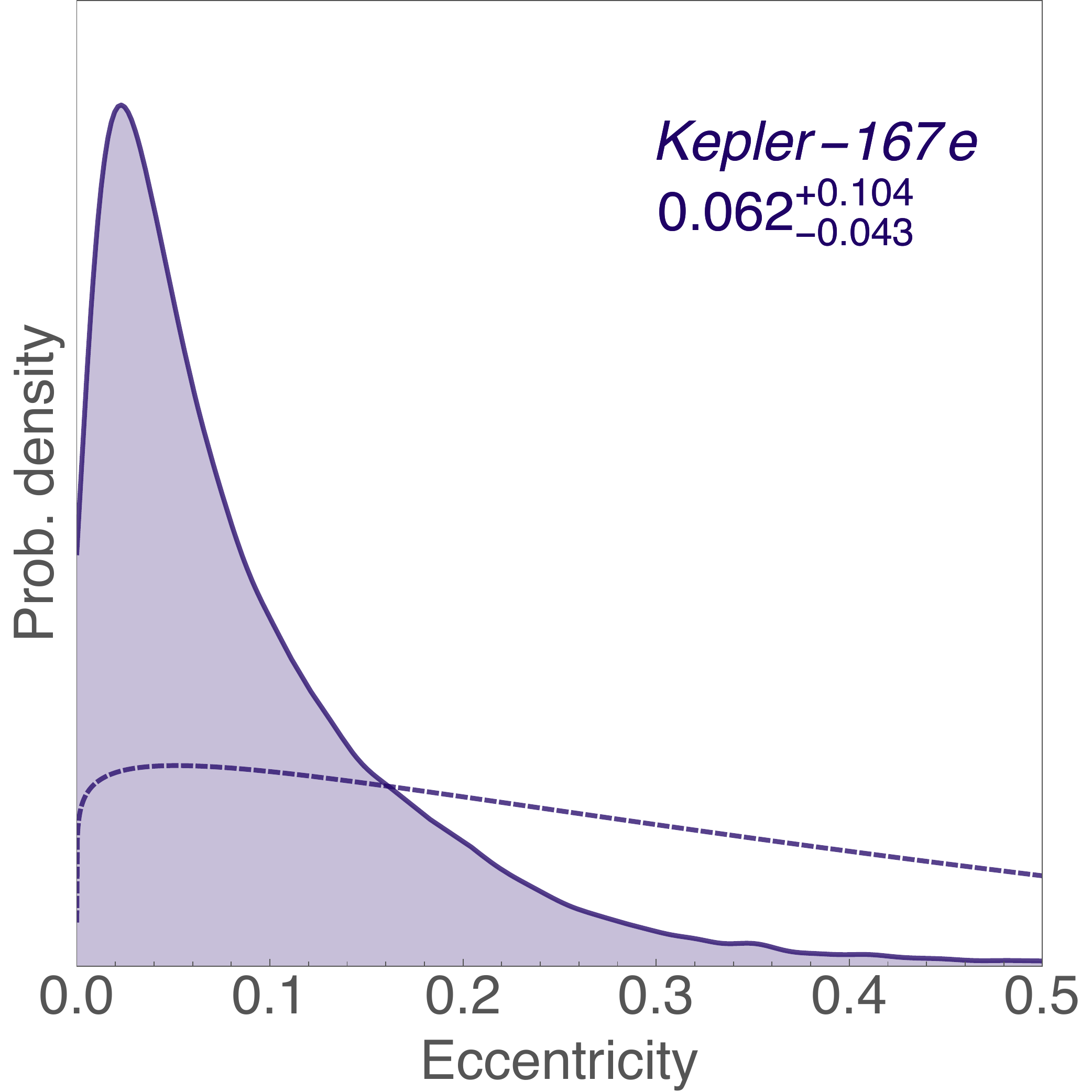}
\caption{
Posterior distribution (solid) of the orbital eccentricity of \kepe,
with comparison to the prior (dashed) describing long-period transiting
planets \citep{eprior:2014}. The eccentricity is constrained by the
comparison of the transit light curve shape of \kepe\ to the $\rho_{\star}$
constraint from the earlier joint fit of \kepb, c \& d.
}
\label{fig:eccentricity_e}
\end{center}
\end{figure}

Unlike the case of \kepd, the light curve plus stellar density prior
does provide a more constraining posterior on the derived eccentricity
than the prior. As shown in Figure~\ref{fig:eccentricity_e}, the orbit
is measured to be close to circular, with $e=0.062_{-0.043}^{+0.104}$.
Using the posteriors for the fundamental stellar parameters derived
using SPC along with the \kepb, c \& d $\rho_{\star}$ constraint plus
isochrone matching, we estimate that \kepe\ is $\sim10$\% smaller
than Jupiter at $10.15_{-0.23}^{+0.24}$\,$\rearth$.

\begin{figure*}
\begin{center}
\includegraphics[width=18.0 cm]{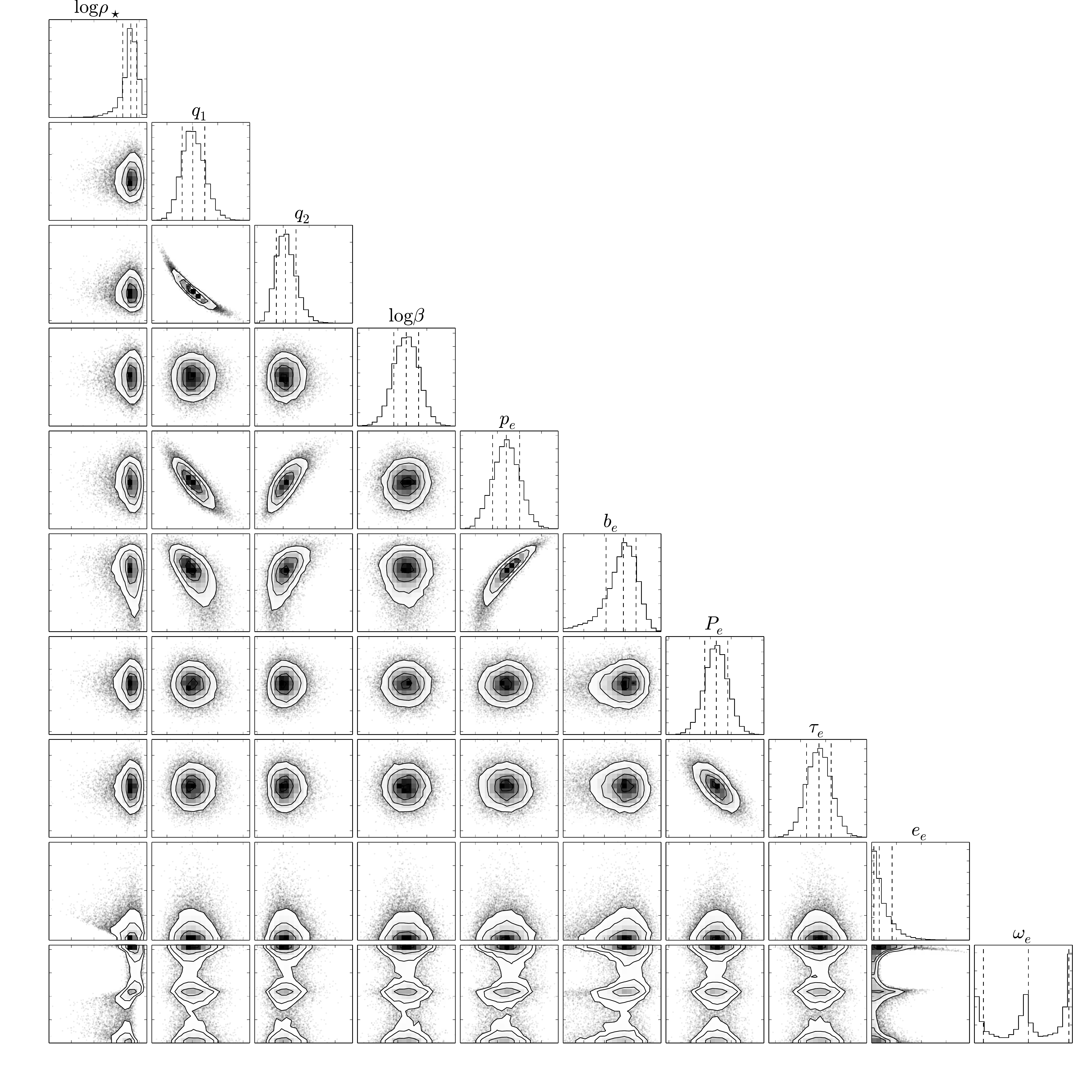}
\caption{
Triangle plot of the posterior distributions of the 10 parameters explored
in the fit of \kepe. Contours mark the 0.5, 1.0, 1.5 \&
2.0\,$\sigma$ confidence intervals and dashed lines on the histograms mark
the median and surrounding 1\,$\sigma$ confidence interval. Posteriors may
be downloaded at \wwwcoolworlds.
}
\label{fig:K167e_post}
\end{center}
\end{figure*}

The posteriors from the fit of \kepe\ are shown in Figure~\ref{fig:K167e_post}
and are available at \wwwcoolworlds. The median and 68.3\% credible intervals
for the basic parameters of \kepb, c, d \& e are shown in Table~\ref{tab:finalparams}.

\begin{table*}
\caption{Final parameter estimates for the planets orbiting \kep.
$^{\ddagger}$ = fixed;
$^{\dagger}$ = assuming a Bond albedo similar to Jupiter of 0.34 (whereas we simply
adopt 0 for the other cases due to the uncertainty in what kind of planet they
are, as shown in Figure~\ref{fig:forecast}).
$^{*}$ = equivalent semi-major axis of the planet if it orbited the Sun
with $e=0$ and insolation level $S_{\mathrm{eff}}$.
} % title of Table
\centering % used for centering table
\begin{tabular}{c c c c c} % centered columns (7 columns)
\hline\hline
Parameter & \kepb & \kepc & \kepd & \kepe \\ [0.5ex] % inserts table
%heading
\hline
% FITTED
\textit{Fitted parameters} & & \\
\hline
$\log_{10}[\rho_{\star}$\,(kg\,m$^{-3}$)] \dotfill &
$3.460_{-0.065}^{+0.031}$ & $3.460_{-0.065}^{+0.031}$ & $3.460_{-0.065}^{+0.031}$ &
$3.460_{-0.065}^{+0.031}$ \\
$q_1$ \dotfill &
$0.63_{-0.30}^{+0.26}$ & $0.63_{-0.30}^{+0.26}$ & $0.63_{-0.30}^{+0.26}$ &
$0.452_{-0.063}^{+0.072}$ \\
$q_2$ \dotfill &
$0.17_{-0.10}^{+0.18}$ & $0.17_{-0.10}^{+0.18}$ & $0.17_{-0.10}^{+0.18}$ &
$0.463_{-0.053}^{+0.062}$ \\
$\log_{10}\beta$ \dotfill &
$-1.976\pm0.036$ & $-1.976\pm0.036$ & $-1.976\pm0.036$ &
$-1.976\pm0.036$ \\
$(R_P/R_{\star})$ \dotfill &
$0.02036_{-0.00032}^{+0.00034}$ & $0.01952_{-0.00044}^{+0.00042}$ & $0.01507_{-0.00052}^{+0.00050}$ &
$0.12810_{-0.00093}^{+0.00091}$ \\
$b$ \dotfill &
$0.17_{-0.12}^{+0.18}$ & $0.25_{-0.13}^{+0.13}$ & $0.474_{-0.063}^{+0.076}$ &
$0.233_{-0.068}^{+0.049}$ \\
$P$ [days] \dotfill &
$4.3931632_{-0.0000045}^{+0.0000046}$ & $7.406114_{-0.000011}^{+0.000012}$ & $21.803855_{-0.000119}^{+0.000078}$ & $1071.23228_{-0.00056}^{+0.00056}$ \\
$\tau$ [BKJD$_{\mathrm{UTC}}$-2,455,000] &
$831.78317_{-0.00036}^{+0.00034}$ & $552.15774_{-0.00084}^{+0.00129}$ & $669.7934_{-0.0018}^{+0.0015}$ & $253.28699_{-0.00040}^{+0.00039}$ \\
$e$ \dotfill &
$0^{\ddagger}$ & $0^{\ddagger}$ & $<0.12$ &
$0.062_{-0.043}^{+0.104}$ \\
$\omega$\,[rads] \dotfill &
- & - & - &
$3.5_{-2.9}^{+2.6}$ \\
\hline
% MODEL INDEPENDENT
\textit{Other transit parameters} & & \\
\hline
$(a/R_{\star})$ \dotfill &
$14.33_{-0.69}^{+0.35}$ & $20.30_{-0.98}^{+0.49}$ & $41.7_{-2.0}^{+1.0}$ &
$560_{-15}^{+11}$ \\
$u_1$ \dotfill &
$0.63_{-0.20}^{+0.15}$ & $0.63_{-0.20}^{+0.15}$ & $0.63_{-0.20}^{+0.15}$ &
$0.915_{-0.019}^{+0.020}$ \\
$u_2$ \dotfill &
$0.14_{-0.25}^{+0.28}$ & $0.14_{-0.25}^{+0.28}$ & $0.14_{-0.25}^{+0.28}$ &
$-0.243_{-0.038}^{+0.040}$ \\
$i$\,[$^{\circ}$] \dotfill &
$89.33_{-0.80}^{+0.47}$ & $89.30_{-0.43}^{+0.36}$ & $89.352_{-0.140}^{+0.090}$ &
$89.9760_{-0.0052}^{+0.0070}$ \\
$T_{14}$\,[hours] \dotfill &
$2.350_{-0.035}^{+0.035}$ & $2.746_{-0.061}^{+0.096}$ & $3.582_{-0.073}^{+0.131}$ &
$16.13_{-0.34}^{+0.44}$ \\
$T_{23}$\,[hours] \dotfill &
$2.249_{-0.033}^{+0.035}$ & $2.630_{-0.064}^{+0.098}$ & $3.440_{-0.077}^{+0.137}$ &
$12.29_{-0.33}^{+0.38}$ \\
$T_{12} \simeq T_{34}$\,[mins] \dotfill &
$2.89_{-0.086}^{+0.342}$ & $3.36_{-0.14}^{+0.37}$ & $4.09_{-0.27}^{+0.49}$ &
$115.9_{-3.7}^{+3.7}$ \\
\hline
% PHYSICALS
\textit{Physical parameters} & & \\
\hline
$R_P$\,[$R_{\oplus}$] \dotfill &
$1.615_{-0.043}^{+0.047}$ & $1.548_{-0.048}^{+0.050}$ & $1.194_{-0.048}^{+0.049}$ &
$10.15_{-0.23}^{+0.24}$ \\
$a$\,[AU] \dotfill &
$0.0483_{-0.0025}^{+0.0017}$ & $0.0684_{-0.0035}^{+0.024}$ & $0.1405_{-0.0071}^{+0.0050}$ &
$1.890_{-0.067}^{+0.058}$ \\
$T_{\mathrm{eq}}$\,[K] \dotfill & $914_{-16}^{+26}$ & $768_{-14}^{+21}$ & $536.0_{-9.6}^{+14.4}$ & $130.9_{-3.0}^{+2.0}$ $^{\dagger}$ \\
$S_{\mathrm{eff}}$\,[$S_{\oplus}$] \dotfill & $115.8_{-8.0}^{+13.0}$ & $57.7_{-4.0}^{+6.5}$ & $13.68_{-0.95}^{+1.54}$ & $0.0739_{-0.0091}^{+0.0047}$ \\
$a_{\mathrm{eff}}^{*}$\,[AU] \dotfill & $0.0929_{-0.0048}^{+0.0034}$ & $0.1316_{-0.0068}^{+0.0048}$ & $0.270_{-0.014}^{+0.010}$ & $3.64_{-0.13}^{+0.12}$ \\
$M_{\mathrm{forecast}}$ (1\,$\sigma$ interval) \dotfill & $2.4\mearth$--$6.3\mearth$ & $2.2\mearth$--$5.8\mearth$ & $1.2\mearth$--$2.9\mearth$ & $0.3\mjup$--$50\mjup$ \\ [1ex]
\hline %inserts single line
\end{tabular}
\label{tab:finalparams} % is used to refer this table in the text
\end{table*}

\section{DISCUSSION}
\label{sec:discussion}

%% DISCUSSION
%%

\subsection{A Transiting Jupiter Analog}

\begin{figure}
\begin{center}
\includegraphics[width=8.4 cm]{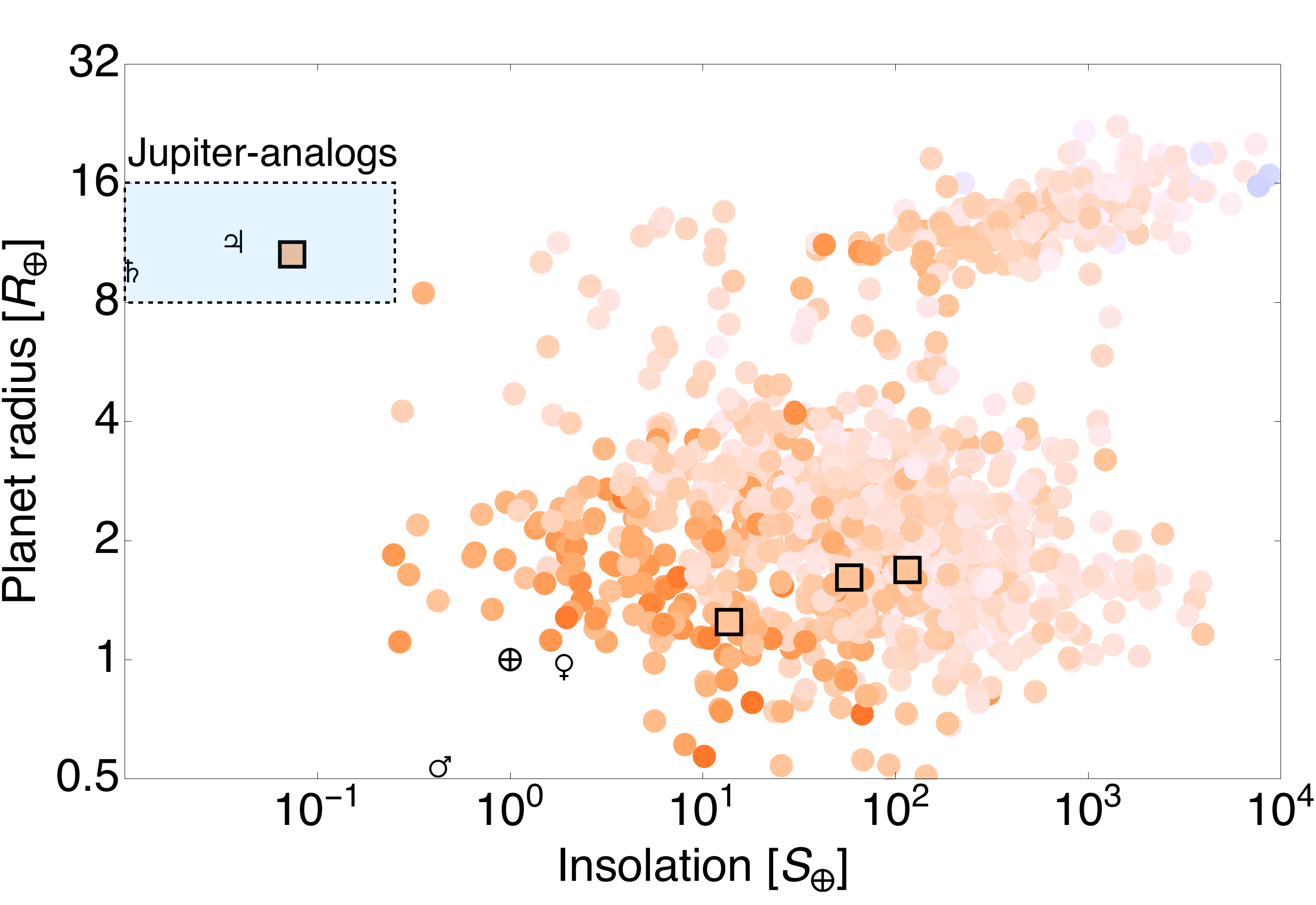}
\caption{
Catalog of known transiting exoplanets with the color depicting the peak wavelength
color of the parent star. Solar System worlds are shown with
black symbols and the \kep\ planets with squares. The blue-box depicts
Jovian-sized planets beyond the snow-line ($\sim0.25$\,$S_{\oplus}$), with
\kepe\ being the first transiting exoplanet to fill this space. Data comes
from the \href{http://www.exoplanets.org}{Exoplanet Orbit Database} \citep{wright:2011}.
}
\label{fig:comparison}
\end{center}
\end{figure}

\kepe\ appears to be the first example of a transiting Jupiter analog,
as defined by its size ($0.91$\,$\rjup$), low eccentricity
($e=0.06_{-0.04}^{+0.10}$) and location beyond the snow-line (see
Figure~\ref{fig:comparison}). Although Jupiter analogs have been found
via other methods (e.g. see \citealt{rowan:2015}), the geometric biases
affecting the transit method make it highly unfavorable for discovering
such worlds.

\kepe\ has an a priori transit probability of $\simeq0.18$\%. Combined with the 
$\eta_{\jupiter}\simeq3$\% occurrence rate of Jupiter analogs \citep{rowan:2015},
one should expect $\mathcal{O}[10]$ transiting examples to exist amongst the
$\sim200,000$ stars observed by \Kepler. However, only those objects just beyond the
snow-line and orbiting later than Solar-type stars will have a chance to produce the
two transits needed to resolve the orbital period. Consequently, it is possible that
\kepe\ could be the only transiting Jupiter analog for which we can precisely measure
the period until the next generation of surveys.

The fact that \kepe\ is transiting offers the opportunity to probe the
atmosphere of a genuine Jupiter analog \textbf{\citep{dalba:2015}}, which has thus far
been impossible. Whilst \kep\ is relatively faint in the $V$ band at 14.3, the fact 
that this is a late-type star means the planet may be characterizable toward the 
near- and mid-infrared bandpasses, where the $K$-band magnitude is 11.8. This fact,
combined with the very deep transit depth of 1.6\%, makes atmospheric characterization
a challenging, but not impossible, task.

As was done by \citet{kipping:2014}, we used the \citet{kennedy:2008} predictions
for the time-evolving snow-line of a $\sim0.8$\,$\msun$ star to estimate that
\kepe's present location corresponds to the snow-line after $\sim800,000$\,years.
This time is less than the median lifetimes of protoplanetary disks of
Solar-type stars \citep[e.g.,][]{strom:1993,haisch:2001} and since disk lifetimes
scale as $M_{\star}^{-1/2}$ \citep{yasui:2012}, \kepe\ could have formed at its
present location. Unlike Kepler-421b \citep{kipping:2014}, which was ``near''
the snow-line, \kepe\ is comfortably beyond it for the majority of the disk
lifetime.\\

\subsection{Could \kepe\ be a brown dwarf?}

With a radius close to that of Jupiter, \kepe\ is a member of the so-called
``degenerate worlds'' class\footnote{Formally, using the \citet{chen:2016}
model, we estimate an 80\% probability that \kepe\ is a degenerate world.},
where the mass-radius relation is nearly flat \citep{chen:2016}. This class
encompasses a diverse range of masses, from that of Saturn to the most massive
brown dwarfs. On this basis, \kepe's radius is consistent with being either a
brown dwarf or a Jovian-like planet. As argued by \citet{chen:2016}, the
division between brown dwarfs and gas giants is somewhat contrived, with both
belonging to a continuum. Nevertheless, we evaluate the possibility that
\kepe's mass is greater than the 13\,$\mjup$ canonical threshold
\citep{spiegel:2011} here.

Using the radius-to-mass probabilistic forecasting code of \citet{chen:2016},
our radius samples can be converted to predicted masses. Being a degenerate
world, this unsurprisingly returns a very broad distribution, with
the 68.3\% credible interval spanning 0.3\,$\mjup$ to 50\,$\mjup$. We find
that the probability of the mass being less than the hydrogen-burning limit
to be 97.4\%, indicating that \kepe\ is very unlikely to be a star.
Moreover, we find 1.33 times more samples below the 13\,$\mjup$ threshold than
above it, implying a slight preference for a Jupiter-like object.

This estimate can be improved by including a suitable occurrence rate
prior, for which we here use \citet{cumming:2008} power-law of occurrence rate
$\sim M^{-0.31}$. This adds further weight to the Jupiter-like scenario with
an odds ratio of 4.28. We therefore estimate that \kepe\ is four times more
likely to be a Jupiter-like planet than a brown dwarf. 

It may be possible for observers exclude the brown dwarf hypothesis
using radial velocities. If \kepe\ is a brown dwarf, then the radial velocity amplitude
would be $K \geq 316_{-11}^{19}$\,m\,s$^{-1}$. In contrast, the (broad) forecasted
radial velocity amplitude is
$\log_{10}[K\,(\mathrm{m}\,\mathrm{s}^{-1})] = 2.32_{-1.47}^{+0.75}$ 
(i.e. $\sim211$\,m\,s$^{-1}$).

\subsection{Multibody Asterodensity Profiling}

The eccentricity of \kepe\ is measured purely using the transit shapes of
the orbiting planets, representing a first for the field. In all
previous cases, independent information constraining the mean stellar
density was used, such as spectroscopic + isochrone analysis
\citep{dawson:2012}, asteroseismology \citep{sliski:2014} or flicker
\citep{kipping:2014}.

The eccentricity of a transiting planet can be measured using asterodensity
profiling \citep{AP:2014}, specifically via the photo-eccentric effect
\citep{dawson:2012}. This essentially compares the light curve derived
stellar density (related to the $T_{14}$ and $T_{23}$ transit durations)
to that derived via some independent method. Although eccentricity is the
dominant effect, for \kep\ the photo-blend effect is in play too,
due to the AO detected companion.

Whilst the most common and accessible method to get an independent
mean stellar density is spectroscopy combined with isochrone modeling,
this approach essentially makes the unrealistic assumption of zero-model
error. Multibody Asterodensity Profiling \citep{MAP:2012} was conceived
with the idea of comparing the light curves of planets orbiting the same
star against one another, to obviate the need to ever go through
evolutionary models. In the case of multiple eccentric planets, the
inverse problem is quite challenging but the compact, inner three planets
of \kep\ are likely on near-circular orbits, providing a so-called 
``stellar anchor'' we can use to characterize the star. This inference is
then used to measure the eccentricity of the outer planet, which a priori
could be much more eccentric.

We are able to measure the eccentricity to be $e=0.06_{-0.04}^{+0.10}$,
which further supports the case that \kepe\ is a Jupiter analog. Critically,
we emphasize that this measurement used nothing more than the \kepler\
photometric time series of a $14^{\mathrm{th}}$ magnitude star and a single
night of AO imaging on a 10\,m telescope. As a comparison, HD\,32963b is a
recently discovered Jupiter analog found using radial velocities
\citep{rowan:2015} for which 199 nights of precise radial velocities on a 10\,m
class telescope led to the comparable constraint of $e=(0.07\pm0.04)$, in spite of
the fact that HD\,32963 is six and a half magnitudes brighter than \kep. However,
the transit method does have the major drawback that transiting Jupiter analogs
are far less numerous than their non-transiting counterparts.\\

\subsection{System Architecture}

\begin{figure}
\begin{center}
\includegraphics[width=8.4 cm]{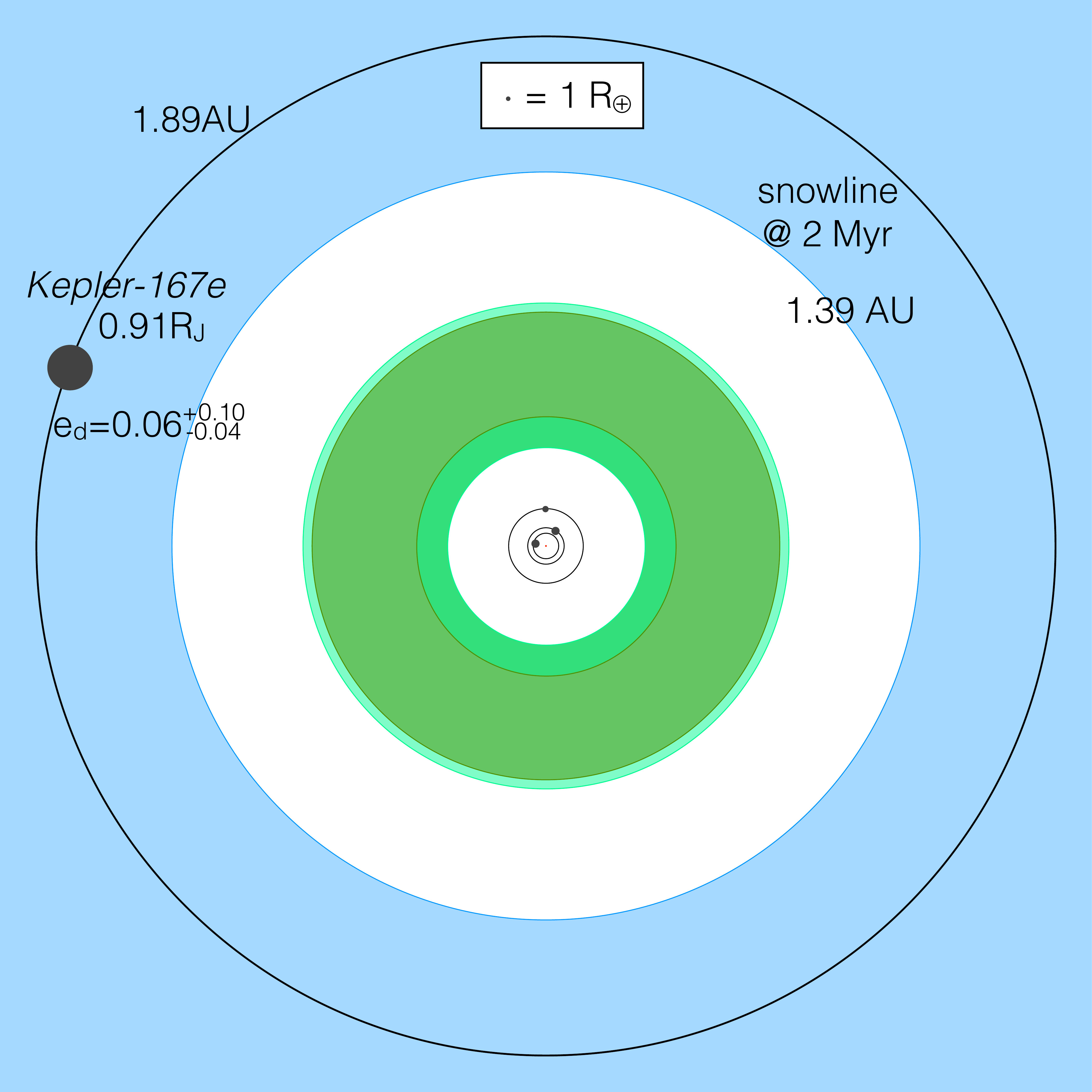}
\caption{
Schematic illustrating the scale of the \kep\ system. Planet sizes are scaled
relative to the key, rather than the orbital distances in order to make
them visible. The four known planets display remarkable coplanarity and
near-circular orbits with the habitable-zone \citep{kopparapu:2013} notably
devoid of transiting planets.
}
\label{fig:schematic}
\end{center}
\end{figure}

All three inner planets orbit interior to the inner edge of the habitable-zone
\citep{kopparapu:2013} and are unlikely to be interesting from an astrobiological
perspective. The architecture of the \kep\ system is curious, with a compact multi
followed by a large cavity of transiting planets and then an outer Jupiter
analog (see Figure~\ref{fig:schematic}). Thus, \kep\ resembles a fusion of the 
compact \kepler\ multis and the classic Solar System. Since transit surveys have a 
very poor sensitivity to long-period planets like \kepe\ (planet yield $\sim P^{-5/3}$;
\citealt{beatty:2008}), it is plausible that \kepe-like planets are found frequently
in the \kepler\ compact multis. A radial velocity survey targeting the bright \kepler\
multis would be able to resolve this question.

\begin{figure}
\begin{center}
\includegraphics[width=8.0 cm]{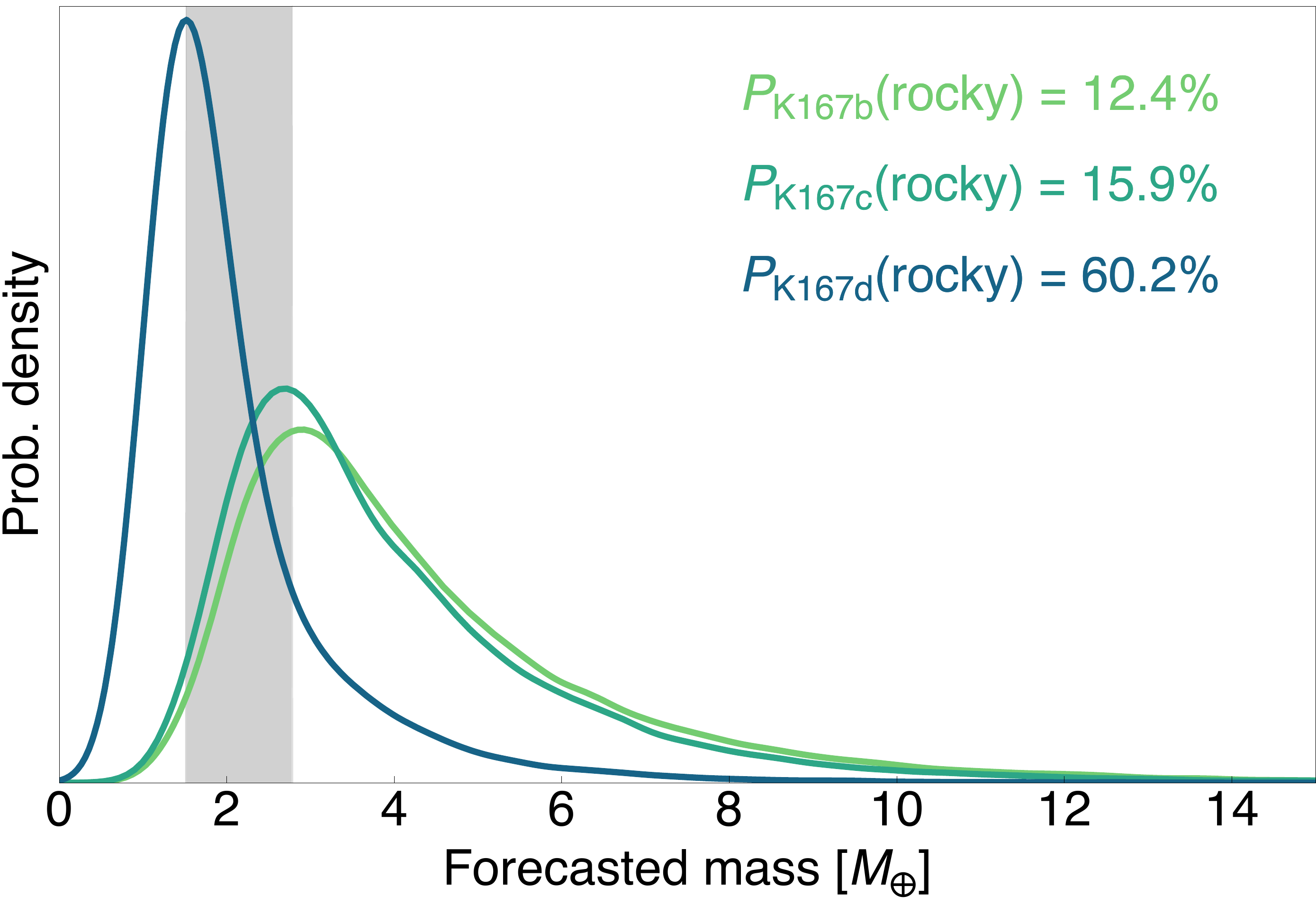}
\caption{
Forecasted masses for the planets \kepb\ (green), \kepc\ (turquoise) and
\kepd\ (blue) using our radii posterior samples and the radius-to-mass
forecasting model of \citet{chen:2016}. Gray region denotes the 1\,$\sigma$
confidence interval of the transition from rocky to gaseous worlds found
by \citet{chen:2016}.
}
\label{fig:forecast}
\end{center}
\end{figure}

Additional non-transiting planets could reside in the \kep\ cavity, although the
known four planets display remarkable coplanarity and low eccentricities,
suggestive of a dynamically cold system. Amongst the inner planets, the planet sizes
increase as one approaches the parent star. Using the \citet{chen:2016} mass-radius
model, we estimate that the inner two planets are most likely gaseous worlds
whilst the outer planet is most likely rocky (see Figure~\ref{fig:forecast}). Whilst
this pattern ostensibly jars our anthropocentric prior, as well as the expected outcome
of photo-evaporation \citep{lopez:2013}, \citet{ciardi:2013} find that there is no
preferential ordering of compact \kepler\ multis for planets $R \lesssim 3 \rearth$.

The \kep\ system teases the possibility that compact multis may plausibly harbor
distant Jupiter analogs, inviting the community to pursue this question with
current and future facilities. Moreover, whilst \kep\ is not a bright star,
transiting Jupiter analogs represent an important new class of targets in
the on-going campaigns to characterize the atmospheres of alien worlds. Discovering
members of this population around brighter stars is a challenging but likely
rewarding task for future missions.

% #####################################################################
%% Acknowledgements
\acknowledgements
\section*{Acknowledgements}

% Dodds thanks
This work made use of the Michael Dodds Computing Facility and the
Pleiades supercomputer at NASA Ames.
% Personal funding
GT acknowledges partial support for this work from NASA grant
NNX14AB83G (\kepler\ Participating Scientist Program).
DMK acknowledges partial support from NASA grant NNX15AF09G
(NASA ADAP Program).
% Exoplanets.org
This research has made use of the Exoplanet Orbit Database
and the Exoplanet Data Explorer at 
\href{http://www.exoplanets.org}{exoplanets.org},
% Corner plot
and the {\tt corner.py} code by Dan
Foreman-Mackey at 
\href{http://github.com/dfm/corner.py}{github.com/dfm/corner.py}.
% Kepler acknowledgement
We offer our thanks and praise to the extraordinary scientists, engineers
and individuals who have made the \emph{Kepler Mission} possible.
% Mauna Kea thanks
Finally, the authors wish to extend special thanks to those of Hawai‘ian
ancestry on whose sacred mountain of Mauna Kea we are privileged to be 
guests. Without their generous hospitality, the Keck observations presented 
herein would not have been possible.
%
%% EOF Acknowledgements

% #####################################################################
%% Bibliography

\section*{Appendix:Posterior Distributions}
\label{app:posteriors}

\end{document}